\documentclass[fleqn,usenatbib,useAMS,usenatbib]{mnras}

\usepackage{graphicx}	
\usepackage{amsmath}	
\usepackage{amssymb}	
\usepackage{multicol}   
\usepackage{bm}		
\usepackage{pdflscape}	
\usepackage{graphicx}
\usepackage{listings}
\usepackage{hyperref}
\usepackage{multirow}
\newcommand{\Msun}{\,M$_{\odot}$}
\newcommand{\Msuny}{\,M$_\odot$\,yr$^{-1}$}

\newcommand{\kms}{\,km\,s$^{-1}$} 

\usepackage[T1]{fontenc}
\usepackage{ae,aecompl}
\usepackage{txfonts}

\title[Gas accretion in Milky Way-like galaxies]{Gas accretion in Milky Way-like galaxies: temporal and radial dependencies}

\author[Nuza et al.]{Sebasti\'an E. Nuza$^{1,2,3}$\thanks{E-mail: snuza@iafe.uba.ar}, 
Cecilia Scannapieco$^{2,3}$, Cristina Chiappini$^{3}$, Thiago C. Junqueira$^{3,4}$, 
\newauthor Ivan Minchev$^{3}$ and Marie Martig$^{5}$
\\
\\
$^{1}$Instituto de Astronom\'{\i}a y F\'{\i}sica del Espacio (IAFE, CONICET-UBA), CC 67, Suc. 28, 1428 Buenos Aires, Argentina\\
$^{2}$Facultad de Ciencias Exactas y Naturales (FCEyN), Universidad de Buenos Aires (UBA), Buenos Aires, Argentina\\
$^{3}$Leibniz-Institut f\"ur Astrophysik Potsdam, An der Sternwarte 16, 14482 Potsdam, Germany\\
$^{4}$Funda\c{c}\~ao CAPES, Setor Banc\'ario Norte (SBN), Quadra 2, Bloco L, Lote 06, Edif\'{\i}cio CAPES, Bras\'{\i}lia, DF, Brasil\\
$^{5}$Astrophysics Research Institute, Liverpool John Moores University, 146 Brownlow Hill, Liverpool L3, 5RF, UK\\
}

\begin{document}
\label{firstpage}
\pagerange{\pageref{firstpage}--\pageref{lastpage}}
\maketitle

\begin{abstract}
One of the fundamental assumptions of chemical evolution models (CEMs) of the Milky Way 
(MW) and other spirals is that higher gas accretion rates are expected 
in the past, and in the inner regions of the Galaxy. This leads to the so-called 
``inside-out disc formation scenario''. Yet, these are probably the most unconstrained 
inputs of such models. In the present paper, we aim at investigating these 
main assumptions by studying how gas is accreted in four simulated MW-like galaxies 
assembled within the $\Lambda$CDM scenario. The galaxies were obtained using two 
different simulation techniques, cosmological setups and initial conditions. 
Two of them are MW candidates corresponding to the chemodynamical model of 
\cite{Minchev2013,Minchev2014} (known as MCM) and the Local 
Group cosmological simulation of \cite{Nuza2014}.
We investigate vertical and radial \emph{gas} accretion on to galaxy discs as a function of cosmic time and disc radius. 
We find that accretion in the MW-like galaxies seem to happen in two distinct phases, 
namely: an early, more violent period; followed by a subsequent, slowly declining phase. 
Our simulations seem to give support to the assumption that the amount of gas incorporated 
into the MW disc exponentially decreases with time, leading to 
current net accretion rates of $0.6-1\,$\Msuny. 
In particular, accretion timescales on to the simulated 
thin-disc-like structures are within $\sim5-7\,$Gyr, consistent with 
expectations from CEMs. Moreover, our simulated MW discs are assembled from the 
inside-out with gas in the inner disc regions accreted in shorter timescales 
than in external ones, in qualitative agreement with CEMs of the Galaxy. 
However, this type of growth is not general to all galaxies and it is intimately 
linked to their particular merger and gas accretion history.  
\end{abstract}


\section{Introduction}

Gas accretion is a key ingredient in chemodynamical evolution models (CEMs) aiming at describing 
the assembly history of galaxies through the chemical imprints of stellar populations. 
Together with the star formation (SF) law, the assumption of a gas accretion recipe 
univocally determines the SF evolution of galaxies.  Despite its importance, 
gas accretion is yet one of the \emph{least} constrained inputs of such models. 
Moreover, CEMs of the Milky Way (MW) have shown that a significant number of observables 
are directly affected by assumptions on gas accretion rates and its temporal and radial dependencies. 
In particular, CEMs matching a large number of Galactic observational constraints assume either 
exponential or Gaussian accretion rates 
\citep[e.g.,][]{Matteucci1989,Prantzos1995,Chiappini1997,Portinari1999,Fenner2003,Chiappini2001,Chiappini2009,Molla2016}. 
Another key assumption of CEMs is that the thin-disc of the Galaxy is postulated to form 
{\it from the inside-out}, i.e. the number of newly born stars per unit time has to decrease 
towards disc outskirts. This is usually modelled assuming a shorter infall timescale 
for the inner regions of the disc \citep[e.g.,][]{Chiappini2001,Hou2001,Hou2002}, together 
with a radial dependence of the star formation efficiency\footnote{For the classical chemical
  evolution results on abundance gradients see e.g.,
  \cite{Pagel2009}, \cite{Matteucci2012} and \cite{Stasinska2012}.}. 
These set of assumptions are key to provide CEMs successfullly reproducing 
observed present-day abundance gradients in the Galactic disc, 
as well as photometric properties of galaxies similar to the MW \citep{Boissier1999}.
Some degeneracy with additional model assumptions (e.g., temporal or radial variations in 
the initial mass function, modelling of galactic winds, etc.) may exist, but their inclusion 
does not seem justified given current observational constraints.

More recently, with the advent of asteroseismology, and in particular thanks to the 
{\it CoRoT} satellite mission \citep{Baglin2006}, it has been possible to constrain the time-evolution 
of the abundance gradients in the MW \citep{Anders2017a,Anders2017b}. This key \emph{new} observational 
constraint is well reproduced by the chemodynamical model of \cite{Minchev2013,Minchev2014} 
(hereafter MCM13, MCM14), 
that combines the chemical modelling provided by CEMs with the dynamical information 
of $N$-body simulations. The reason for this agreement can be attributed mainly to the following: 
(i) the inside-out formation of the disc in their \emph{hybrid-model approach} adopts 
the prescriptions of \cite{Chiappini2009} for the MW thin-disc, 
and (ii) the dynamically self-consistent stellar radial migration is taken from a 
simulated galaxy \citep{Martig2012} that contributes to flatten the original 
gradient of the oldest stars.

Further constraints come from large spectroscopic surveys, which have also shown 
present-day chemical abundance gradients to flatten with increasing height from the Galaxy's 
midplane \citep[e.g.,][]{Chen2012,Boeche2013,Boeche2014,Anders2014,Hayden2014,Minchev2015}. 
Again, these rather new constraints are readily explained by the MCM13 and MCM14 
chemodynamical model (see also \citealt{Minchev2016} for more details and comparison with other 
observational constraints). In this respect, the variation of SF histories as a function of 
cosmic time and galactocentric distance is essential to reproduce chemical abundance 
diagnostic diagrams as, for instance, the classical [$\alpha$/Fe] vs. [Fe/H] plot 
\citep[see][and references therein]{Anders2014,Anders2017a}.

\begin{figure*}
\vspace*{-.5cm}
\hspace*{1.2cm}\includegraphics[scale=0.4,angle=0]{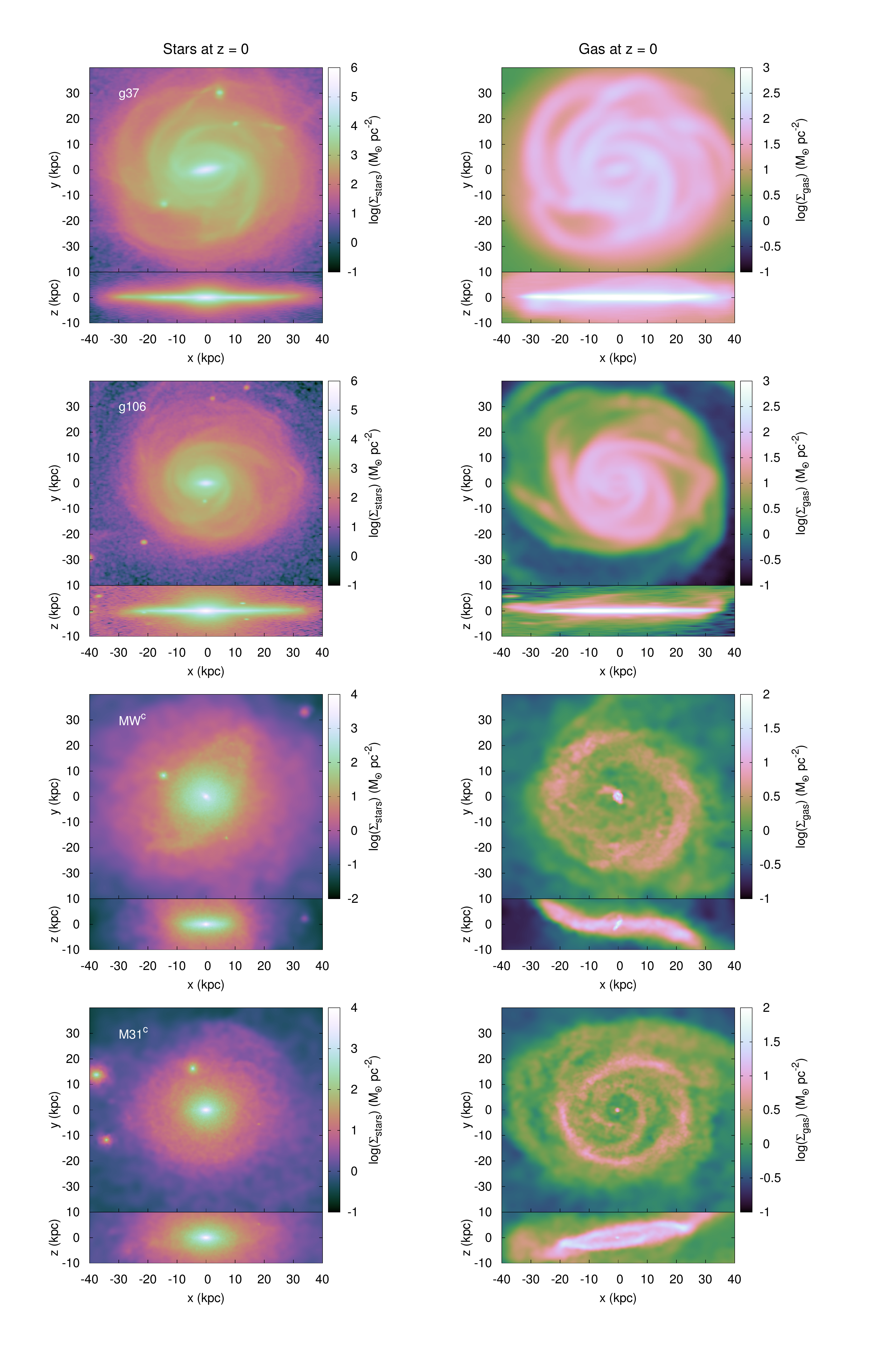}
\caption{Face-on (upper panels) and edge-on (lower panels) density maps of the stellar (left) and gaseous (right) discs 
for the four simulated galaxies considered in this work.}
\label{dens}
\end{figure*}

Direct evidence of ongoing/recent accretion of external gas from minor mergers
and interactions, as well as from halo gas and the local environment
in nearby galaxies --including the MW and Andromeda (M31)-- has been found
\citep[e.g.,][]{Thilker2004,Sancisi2008,Lehner2011,Putman2012,Richter2012,Gentile2013,Richter2017}. 
Spiral galaxies in the local universe show extraplanar H\,{\sc i} structures and so-called
high-velocity clouds (HVCs) of neutral material presumably feeding star formation in galaxy discs
at a rate of about $0.1-0.7\;$\Msuny \citep{Sancisi2008,Richter2012,Putman2012}. Furthermore,
the ionized counterpart of HVCs has also been detected in the MW halo at higher rates,
suggesting that these clouds comprise the bulk of gas accretion in MW-type galaxies \citep[][]{Lehner2011,Richter2017}. 
In fact, the existence of a gas-reservoir around galaxy-size haloes has long been suggested 
by numerical simulations\footnote{See \cite{Nuza2014} for a thorough comparison between
  simulations and observations of the gaseous halo components in MW and M31.}
and it is hinted by observations of the MW and nearby spirals
\citep[e.g.,][]{Thilker2004,Sancisi2008,Miller2015,Richter2017},
and of more massive and distant galaxies \citep[e.g.,][]{Tumlinson2013,Singh2018}.

Other works place constraints based on more indirect methods. For instance,
\citet{Rocha2000} proposed that the empirical age-metallicity relation 
and the SF history (SFH) of the MW allow an estimate of the time variation of gas mass, 
which could lead to an estimate of gas accretion evolution.
\cite{Fraternali2012} used a model-dependent method (excluding chemical information)
to derive gas infall rates in spiral galaxies as a function of 
time and galactocentric distance, applied to 21 galaxies from the THINGS survey 
\citep{Walter2008} and the MW. For the latter, the authors find a general good agreement 
with conclusions coming from pure CEMs.

Nevertheless, a precise derivation of past accretion or ejection rates in galaxies 
is extremely difficult to obtain, either from observations or first principles, calling for the 
use of hydrodynamical simulations in a cosmological framework with enough resolution and 
reliable feedback modelling (important to constrain the role of outflows). 
Current zoom-in, high-resolution, cosmological simulations of MW-like systems are broadly 
consistent with a large number of chemical, dynamical and structural properties of galaxies 
\citep[e.g.,][]{Guedes2011,Dubois2013,Stinson2013,Aumer2014,Genel2014,Hopkins2014, 
Marinacci2014,Naab2014,Wetzel2016,Grand2017}. However, the relative contribution of the different 
feedback processes shaping them 
(e.g., supernova vs. black hole vs. radiation pressure vs. cosmic ray feedback) is still 
under debate \citep[][]{Scannapieco2012}. For MW-mass systems, the release of energy and 
chemical elements by supernovae plays a crucial role in their formation and evolution, enriching 
the interstellar medium and circulating material between the thin 
disc and galactic halo components \citep{Scannapieco2008}. Although recycled gas seems to be underdominant 
in comparison to gas directly accreted from the intergalactic medium, this has not been yet 
studied in detail and might have consequences for CEMs.

Using the chemical evolution approach, \cite{Chiappini1997,Chiappini2001} demonstrated 
that a model assuming two main accretion episodes for the formation of the Galaxy gives a 
reasonable description of observations both in the solar neighborhood and the whole disc. 
According to this model, during first infall, primordial gas is rapidly accreted 
giving rise to the formation of spheroidal components (i.e., stellar halo and bulge)
of the Galaxy. This episode is followed by the smooth accretion
of fresh gas in much longer timescales resulting in the inside-out formation of the thin-disc component.
Recently, by analysing a sample of MW-like systems in the 
EAGLE cosmological simulation, \cite{Mackereth2018} found that a `two-infall' model of the 
kind described above may provide a plausible explanation for the bimodality observed in 
[$\alpha$/Fe] vs. [Fe/H] distribution, although the usual approximation of instantaneous 
mixing of star-forming gas in CEMs can lead to some discrepancies.

\begin{figure}
  \begin{center}
  \includegraphics[width=85mm, height=45mm,angle=0]{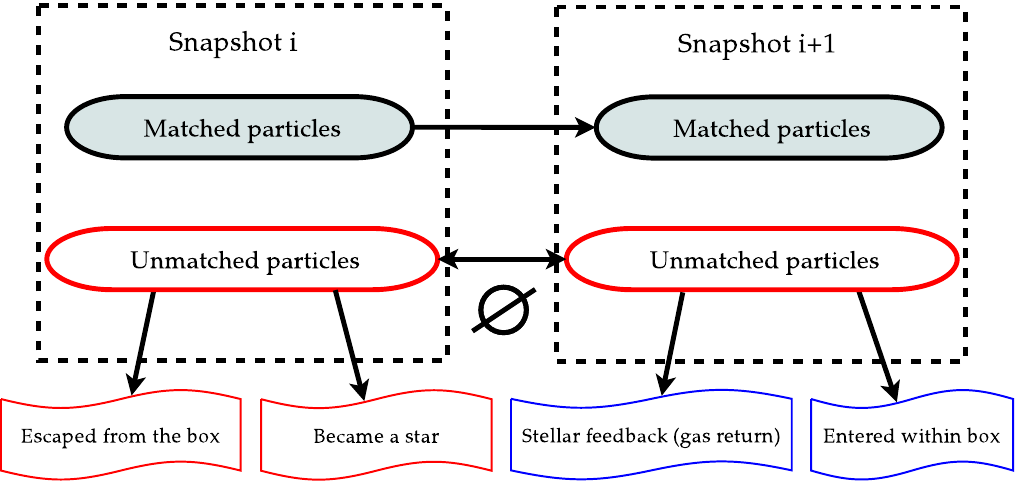}
  \end{center}
  \caption{Flow chart illustrating the matching of particles between snapshots. We cross-checked particles within two consecutive simulation outputs corresponding to times 
  $t$ and $t + \Delta t$, where $\Delta t$ is the timestep. Three cases are considered: {\it i)} particles belonging to both files (matched); 
  {\it ii)} those in snapshot $i$ but not in $i+1$ (unmatched), and {\it iii)} those in snapshot $i+1$ but not in $i$ (unmatched). For the unmatched particles, several 
  possibilities for their origin are indicated.}
  \label{fig_snaps}
\end{figure}

In this work, we compute gas flows on to galactic discs of a set of 
simulated MW-like systems and study their gas accretion history in 
relation to the usual assumptions of CEMs\footnote{Preliminary results 
can be found in \cite{Nuza2018}.}. We consider the exchange of 
gas between the stellar disc and its surroundings, as well as radial 
migration within the disc. Our sample consists of 
four MW-like galaxies from \cite{Martig2012} and \cite{Nuza2014} plus 
a rescaled version of one galaxy of the former following the 
procedure of MCM13 and MCM14. 
These galaxies correspond to two different simulation suits, 
each of them with their own advantages and shortcomings, that serve to 
gauge the variance in existing self-consistent galaxy formation 
models. However, the discussion concerning the detailed 
modelling of the physics of accreted gas in simulations is out of the 
scope of this paper\footnote{For a recent discussion on the subject see 
e.g., \cite{Nelson2016} and references therein.}.

The structure of the paper is as follows. In Section~\ref{simu}, we present the 
cosmological simulations used to obtain the MW-like galaxies analysed in this 
work, their properties and the main differences between the two galaxy formation models. 
In Section~\ref{MCF} we describe the method used to compute gas fluxes 
on to the simulated stellar discs. 
Results are presented in Section~\ref{res}, where we focus on gas fluxes 
and accretion patterns vs. cosmic time and radius for material coming 
from directions perpendicular and parallel to the disc plane. 
Finally, in Section~\ref{concl} we summarise and discuss our results.

\section{Cosmological simulations}
\label{simu}

We analyse four galaxies labeled g37, g106, MW$^{\rm c}$ and M31$^{\rm c}$, 
which were obtained using two different simulation techniques, cosmological setups 
and initial conditions. Galaxies g37 and g106 are obtained assuming a WMAP-3 cosmology. 
They belong to the \cite{Martig2012} sample of simulated galaxies and were chosen 
to display a quiet merger history. 
On the other hand, galaxies MW$^{\rm c}$ and M31$^{\rm c}$ are the MW and M31 
candidates of \cite{Nuza2014} (see also \citealt{Scannapieco2015}) that 
were extracted from a WMAP-5 cosmological simulation of the Local Group. 
As a result, these galaxies reside in an environment more similar to the actual one. 
The main properties of each model galaxy at $z=0$ are shown in Table~\ref{galaxies}. 
The latter also includes a scaled-down version of g106, referred to as g106r, that 
was used as a MW candidate 
by MCM13 and MCM14 in their Galactic chemodynamical model. 
More details concerning the galaxies, as well as the main assumptions and 
techniques adopted to perform each simulation, can be found below.  

\begin{table*}
\caption{Properties of the simulated galaxies at $z=0$. The columns show virial mass, virial radius, 
stellar mass (within optical radius), optical radius, bulge-to-total ratio and references.}
\label{galaxies}
\begin{tabular}{c c c c c c c}
\hline
Name & $M_{\rm vir}$     & $R_{\rm vir}$ & $M_{*}$           & $R_{\rm opt}$   & B/T & Reference\\      
     & [$10^{10}$ \Msun] & [kpc]         & [$10^{10}$ \Msun] & [kpc] &     &          \\      
\hline

g37       & 120 & 220  & 12.0 & 35.1 & 0.13 & \cite{Martig2014}\\
g106      &  42 & 154 & 4.3 & 24.6 & 0.22 &\cite{Martig2014} \\  
MW$^{\rm c}$      & 125 & 222 & 4.3 & 14.7 & 0.43 & \cite{Nuza2014}\\
M31$^{\rm c}$     & 168 & 245 & 4.7 & 15.3 & 0.73 & \cite{Nuza2014}\\
g106r$^{\dagger}$ & 28 & 92 & 7.9 & 15.0 & 0.22&  MCM13, MCM14\\

\hline
\end{tabular}\\
$^{\dagger}$Rescaled g106 used as a MW proxy by MCM13 and MCM14 (see Section~\ref{MWcandidates}).
\end{table*}

\subsection{g37 and g106}
\label{Martig_gxs}

These galaxies are selected from the larger galaxy sample of \cite{Martig2012}. 
The simulation technique involves basically two steps \citep[see][for technical details]{Martig2009}. 
First, a dark-matter-only simulation is performed in a box of $20\,h^{-1}\,$Mpc on a side using the adaptive 
mesh refinement RAMSES code \citep{Teyssier2002} to extract merger and accretion histories for the target haloes. 
Some properties such as time, mass, velocity and spin of incoming satellites are then stored. 
In the second step, a new simulation is performed in which each halo recorded in the dark-matter-only run 
is replaced by a model galaxy made up of gas, stars and dark matter particles. Similarly, each diffuse particle of 
the cosmological simulation is replaced with a blob of higher-resolution gas and dark matter particles. 
These model `seed' galaxies are made up of a disc containing gas and stars, a stellar bulge and a 
dark matter halo following analytical prescriptions for the profiles in each case.  

The stellar mass of a galaxy in the simulation is set according to the \cite{Moster2010} relation as a 
function of halo mass and redshift. All model galaxies are assumed to be disc galaxies without a bulge, 
except for the most massive haloes with $M_{\rm vir}>10^{11}\,$\Msun\,at $z<1$, in which case the mass of 
the bulge is set to $20\%$ of the stellar mass. The gas content in the disc is also set according to an 
analytical scheme: at $z<1$, the gas fraction (with respect to the total baryonic mass) is 0.3 for small 
galaxies ($M_{\rm vir}<10^{11}\,$\Msun) and 0.15 for the massive ones. At higher redshifts, the gas fraction 
is chosen independently of the total mass, and is set to 0.5 for $1<z<3$ and 0.7 for $z>3$. 
No hot gas halo is included, which should be a sensible approximation only for halo masses lower than a 
critical mass of $\sim$$10^{12}\,$\Msun\,, which is required for virial shocks to be stable 
\citep[e.g.,][]{BirnboimDekel2003}. 

The resimulation starts at $z=5$ inside a region of radius 800\,kpc, i.e. spanning a few times the 
virial radius value of the main galaxy at $z=0$, and follows its evolution down to the present epoch. 
The mass resolution is set to $1.5\times10^{4}\,$\Msun\,for gas, $7.5\times10^{4}\,$\Msun\,for stars 
and $3.5\times10^{5}\,$\Msun\,for dark matter. Note that in these simulations there are no particles 
outside the 800\,kpc sphere. 
Gas dynamics is modelled with a sticky-particle scheme including also the effect of star formation, 
(kinetic) supernova feedback and the continuous mass loss rate from stars (see \citealt{Martig2012} 
and references therein for details on the modelling of these processes).

The target haloes are selected in such a way that they inhabit relatively isolated environments 
at $z=0$, i.e. no halo more massive than half of their mass can be found within $2\,$Mpc and they are, 
at least, at a distance of $6\,$Mpc from one of the four most massive haloes in the simulation. 
Therefore, by construction, the target haloes are located in underdense regions with typical environmental 
density between that of the centre of voids and the LG. As a result, the merger history of the target 
haloes displays a relatively quiet behaviour with none or few significant mergers at late times. 
In particular, g37 shows a quiescent history\footnote{Following \citet{Scannapieco2015}, we refer 
to `minor', `intermediate' or `major' mergers when the {\it total} mass ratio of the colliding 
galaxies is $<5\%$, $5-30\%$ or $>30\%$, respectively.}: all interactions at $z\lesssim1.5$ are `minor', 
displaying stellar mass ratios smaller than 1:50. The situtation is similar for g106 which has only 
one `major' merger at $z\sim0.7$ with a stellar mass ratio of 1:5. Such a quiet merger history promotes 
the formation of galaxies with dominant stellar discs.

The main advantage of this numerical technique is the low computation time needed to perform the 
resimulations. This allows to increase the mass resolution at a lower computational expense. 
Also, the fact that the evolution of the main galaxy is decoupled from the expansion of the universe 
keeps the physical resolution constant as a function of time at no additional cost. The most important 
drawbacks are related to the large number of free parameters needed to model the seed galaxies; 
the ad-hoc distribution of gas within the lagrangian region, and the treatment of hydrodynamics using 
a sticky-particle model that poorly treats the hot gas phase in haloes with masses above the critical 
$\sim$$10^{12}\,$\Msun\,value. This limitation implies that the main accretion mode for g37 and g106 
consists of cold gas coming from neighboring filaments.

\subsection{MW$^{\rm c}$ and M31$^{\rm c}$}
\label{CLUES_gxs}

In our study we also analyse the MW and M31 candidates of \cite{Nuza2014} which have been 
extracted from a simulation of the LG performed within the context of the CLUES 
project\footnote{\url{www.clues-project.org}}. The simulation is run using the Tree-PM Smoothed 
Particle Hydrodynamics GADGET-3 code \citep{Springel2005,Springel2008} including modules for 
star formation, metal-dependent cooling, chemical enrichment, (thermal) supernova feedback, 
a multiphase model for the gas component and a UV background 
field \citep[see e.g.,][and references therein, for details concerning the modelling of these processes]{Scannapieco2015}.

The initial conditions (ICs) reproduce, by construction, the known dynamical properties 
of the local Universe at the present day within a dark-matter-only periodic box 
of $64\,h^{-1}\,$Mpc on a 
side \citep[see][for details on the method and the type of constraints used]{Yepes2014}. 
The simulation starts at $z=50$ inside a {\it zoomed-in} sphere of $2\,h^{-1}\,$Mpc radius, 
located at the center of the box, containing so-called {\it high-resolution} gas and dark 
matter particles with a mass of $5.6\times10^{5}\,$\Msun\,and $2.8\times10^{6}\,$\Msun\,respectively. 
The size of this region is large enough to comprise the large-scale distribution of gas in the 
LG associated to the MW and M31 galaxy candidates during their evolution. Additionally, this also 
ensures that the transition between the {\it low-} and {\it high-resolution} zones occurs away 
from the position of the target galaxies thus decreasing the chances of contamination with 
low-resolved particles in the region of interest. 

As galaxies evolve, the influence of the LG environment on the formation of galaxy candidates 
is taken into account\footnote{The interested reader is also referred to \cite{Creasey2015} which 
focuses on the effect of the LG environment in the final galaxy properties for a related set of runs.}. 
As a result, gas inflows are consistently included in the simulation, eventually giving rise to the 
formation of stellar discs in a natural way. This cosmological setup enables us to test the assembly 
history of a Milky Way-type galaxy without introducing any other simplifying assumptions, 
i.e. no ad-hoc prescriptions for the gas distribution, merger trees and/or `seed' galaxies 
at early times are assumed. As shown by \cite{Nuza2014} both `cold' an `hot' accretion modes are considered 
throughout the evolution with warm-hot material significantly contributing to the accreted mass budget. 
This is clearly the main advantage of full cosmological simulations such as the present one, although, 
in this case, model consistency is achieved at the expense of a lower numerical resolution.

At $z=0$, the target haloes inhabit an environment similar to that of the actual LG. The two main 
galaxies in this region, labelled MW$^{\rm c}$ and M31$^{\rm c}$, are separated by a distance 
of $652\,$kpc and are approaching each other with a radial velocity of $-138\,$km\,s$^{-1}$, 
resembling the actual MW-M31 system. However, it is worth noting that, as the constraints 
in the ICs can only be imposed on large scales, the properties of the simulated MW$^{\rm c}$ and 
M31$^{\rm c}$ will not necessarily reproduce {\it in detail} those of the actual ones.

The more realistic LG scenario in this simulation suggests that the merger history of 
the target galaxies has to be more active than in the g37 and g106 cases. 
In fact, at $z<1$, MW$^{\rm c}$ undergoes four `intermediate' mergers: two at $z=0.81$ 
and $0.03$ with a stellar mass ratio of 1:20; one at $z=0.6$ with a ratio of about 1:10, 
and one at $z=0.29$ with a ratio of 1:50. M31$^{\rm c}$ undergoes two `intermediate' 
mergers at $z=0.82$ and $0.32$ with a stellar mass ratio of 1:20, and two `major' mergers 
at $z=0.28$ and $0.25$ with a stellar mass ratio of about 1:4 and 1:3, respectively 
(see Table~2 of \citealt{Scannapieco2015}). In particular, these two last `major' mergers 
are responsible of significantly reducing the growing stellar disc of the simulated 
M31$^{\rm c}$ leading to a $z=0$ bulge-to-total ratio larger than for the actual M31. 
However, this characteristic is not relevant for our study as here we are only interested 
in studying the evolution of gas accretion onto MW-size haloes regardless of the detailed 
properties of the resulting stellar discs.

The stellar and gas densities for the simulated galaxies at the present time can be seen in 
the left and right columns of Fig.~\ref{dens} respectively, where we show the first four galaxies 
in Table~\ref{galaxies} (from top to bottom). Galaxy g106r is not shown as it is just a scaled-down 
version of g106. Every panel depicts both face- and edge-on views in the reference frame corresponding 
to the stellar or gaseous discs; meaning that their orientation defines the $xy$-plane 
of the systems for each component (see Section~\ref{Box}). As expected, g37 and g106 galaxies display a 
very thin and large stellar disc (see $R_{\rm opt}$ in Table~\ref{galaxies}) which is always aligned 
to the gaseous one at late times. In addition, the bulge counterpart is small. This is not always 
the case for the CLUES galaxies owing to the larger number of interactions undergone by them 
evidenced, for instance, by the presence of warps in the gas distribution. This indicates that 
considering target galaxies in the right environment can play an important role in the final galaxy 
properties.

\section{Computing gas fluxes}
\label{MCF}

The flux of particles per unit time at an arbitrary surface $S$ is defined as

\begin {equation} 
\label{flux_def}
	F = \frac{N}{S \Delta t}, 
\end {equation}

\noindent where $N$ is the number of particles crossing the area $S$ and $\Delta t$ is the 
time interval considered. Ideally, one would like to sample this function in a continuous way. 
However, simulations only provide a discrete set of output files indicating the spatial and 
kinematical information of the gas particles. It is possible to think of these outputs as a 
collection of pictures, or {\it snapshots}, of the simulated system at different stages 
in its evolution. Therefore, in order to compute the flux of gas particles as a function of time 
using Eq.~\ref{flux_def}, we need to know the particle trajectories between the different snapshots. 
In this way, we will be able to compute the approximate crossing time and position of gas particles 
accross any given surface. 

As a first approximation, one could think of estimating particle trajectories using a 
linear path between snapshots if the corresponding time interval is small enough. 
However, this approximation cannot be adopted in our case, mainly because the collisional 
nature of gas can significantly depart particles from a linear trajectory. In addition, 
gas particles may {\it appear} or {\it disappear} from the studied region, e.g. after being 
ejected as a result of supernova feedback or, eventually, transformed into a star particle 
if the conditions for star formation are fulfilled.  

A schematic view of all possible cases between two consecutive output files can be seen 
in Fig.~\ref{fig_snaps}, where gas particles have been divided among those found in both snapshots 
(matched particles) and those without a matched counterpart (unmatched particles). For the sake 
of simplicity, when computing the fluxes, we will ignore unmatched particles as there is no 
obvious way to track their trajectories. Fortunately, this type of particles only represents 
a small fraction of the total number ($\lesssim 1-5\%$, depending on the simulation) for most of 
the evolution; specially after the first accretion episode.

Once we have identified the particles belonging to all consecutive output files, 
we calculate their acceleration using the initial and final positions in order to 
track the trajectories more precisely. This allow us to estimate the spatial and time 
coordinates of all particles crossing the stellar disc boundary by dividing the time interval 
between snapshots into smaller timesteps. In the next section, we will present more details 
concerning the choice of coordinate system and the flux calculation.

\subsection{Defining the surfaces}
\label{Box}

We follow the accretion history of our simulated galaxies adopting, in each case, 
a coordinate system aligned to the total angular momentum of the stellar discs. In this way, 
we are able to follow the orientation of the galaxies as they evolve to determine the amount 
of material flowing across the vertical and radial directions with respect to the plane defined 
by the stellar discs. Therefore, at any given cosmic time, a flat cylinder comprising the 
rotationally supported stellar component in the $z=0$ plane is adopted. 
The cylinder is divided into several rings of radius $r_i$ and width $\Delta r$ 
to study the radial dependence of flux up to a distance $R_{\rm d}$. 
The vertical coordinate $z_i$ of the cylinder goes from 
$-z_{\rm d}$ to $z_{\rm d}$, where $z_{\rm d}$ defines the cylinder's height. 
We note that, at very early times, vertical and radial directions are essentially 
indistinguishable as only weak rotational support is achieved by the stellar component 
at these stages.

\begin{figure}
\includegraphics[scale=0.6,angle=0]{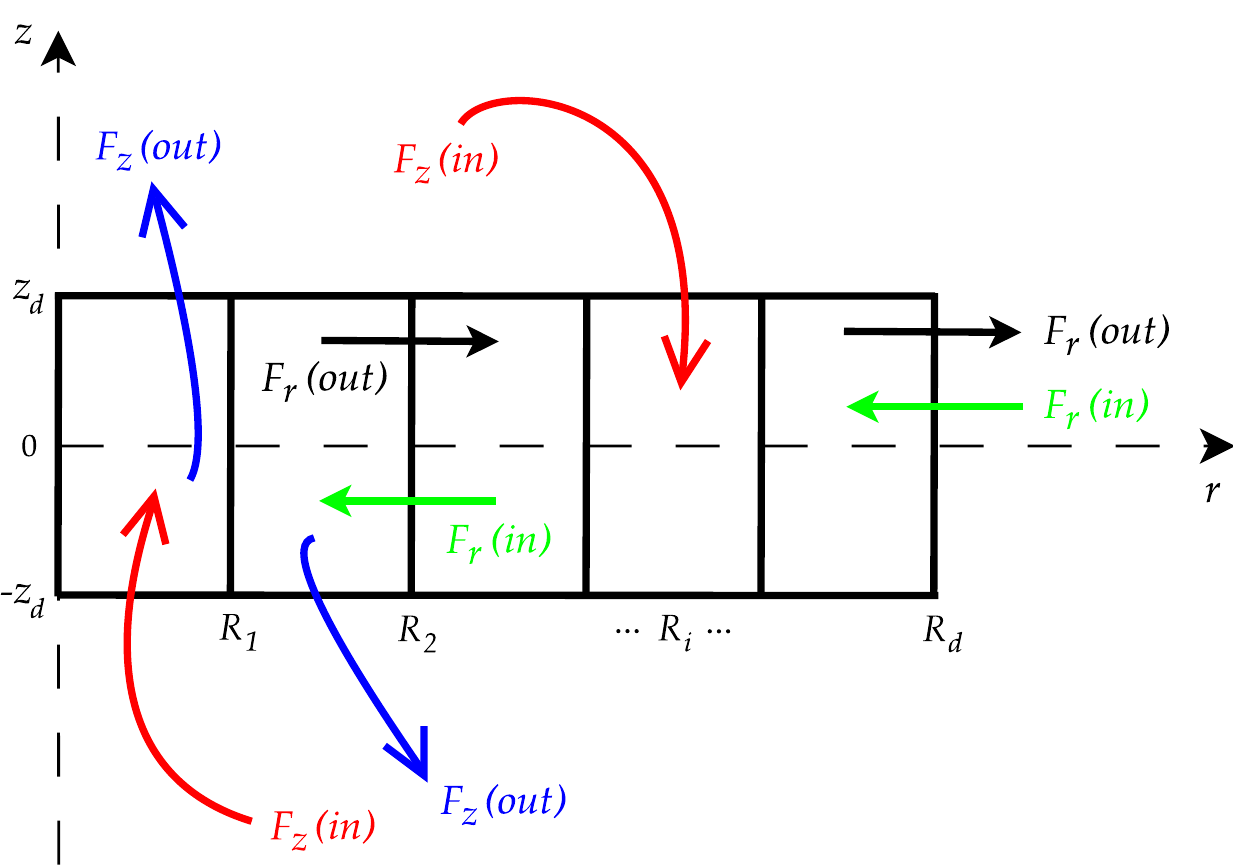}
\caption{Scheme of the gas flux computation on to a stellar disc located 
in the $z=0$ plane. A cylindrical coordinate system comprising the disc is divided 
into rings of width $\Delta r$ up to a distance $R_{\rm d}$. 
The $z$-coordinate goes from -$z_{\rm d}$ to $z_{\rm d}$. 
Red and blue arrows represent inflowing and outflowing {\it vertical} 
fluxes at $|z|=z_{\rm d}$. Similarly, green and black arrows represent 
inward and outward {\it radial} fluxes.}
\label{fig3}
\end{figure}

Fig.~\ref{fig3} shows a scheme depicting the adopted reference frame as well as the 
gas fluxes across the radial and vertical directions; $F_{z}(in)$ and $F_{z}(out)$ 
represent the flux of infalling and outflowing material across the arbitrary surfaces $S_z$, 
whereas $F_{r}(in)$ and $F_{r}(out)$ stand for the flux across the surfaces 
$S_r$, i.e. inwards and outwards towards the galactic centre. 
Thus, using the flux definition given by Eq.~\ref{flux_def}, we can compute the 
time-dependent particle fluxes as function of radius as follows 

\begin {equation} 
\label{flux_z}
F_z(r,t) = \frac{N_z}{S_{z}\,\Delta t}{\rm ;} 
\end {equation}

\begin {equation} 
\label{flux_r}
F_r(r,t) = \frac{N_r}{S_{r}\,\Delta t}{\rm ,}
\end {equation}

\noindent where $N_z$ and $N_r$ are the number of particles crossing the 
surfaces during time $\Delta t$, defined by the timestep between two consecutive 
output files (see Figs.~\ref{fig_snaps} and~\ref{fig3}). 
In what follows, we will convert all fluxes to (physical) units of M$_{\odot}$\,Gyr$^{-1}$\,pc$^{-2}$ 
by transforming the corresponding particle number into mass. If all gas particles have 
the same mass resolution this is a straightforward task; otherwise, we must integrate 
over particle masses as they cross the surfaces of interest. 

\begin{figure*}
\includegraphics[scale=.6,angle=0]{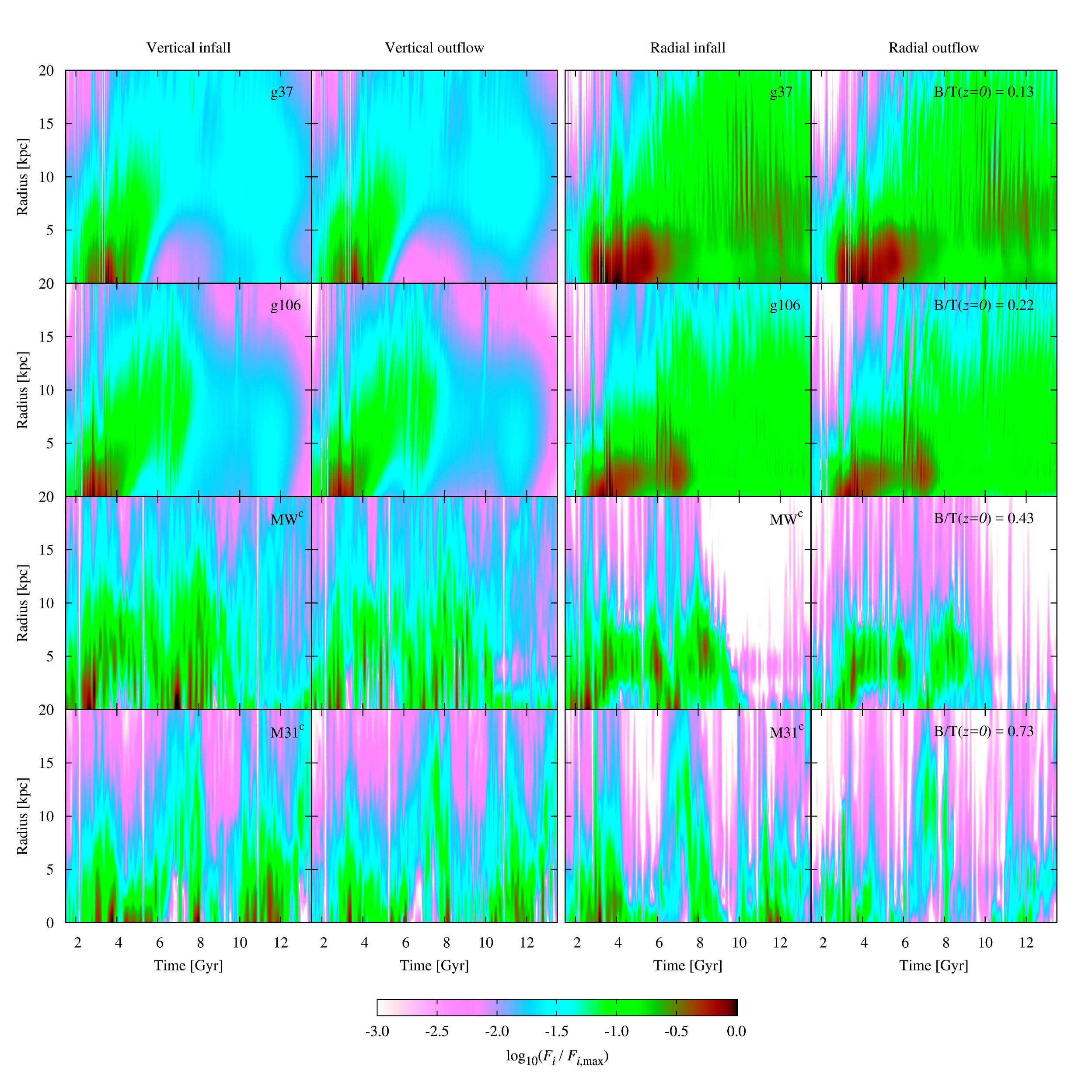}
\caption{Normalized logarithm of gas fluxes as a function of time and radius for the four 
simulated galaxies (from top to bottom). The normalization is performed using the peak 
{\it infall} value of each component. The bulge-to-total stellar mass ratio of each galaxy 
at $z=0$ is also indicated.}
\label{fig4}
\end{figure*}

The amount of mass deposited on an arbitrary surface $S_i$ during a certain period of time 
can be written as  

\begin {equation} 
\label{Ac}
\dot{M_i}(r,t) = F_{m,i}(r,t)\,S_i,
\end {equation}

\noindent where $F_{m,i}$ represents the inflowing/outflowing gas mass flux accross surface $S_i$ 
as computed from Eqs.~\ref{flux_z} and~\ref{flux_r}. 
The integrated mass deposition over the entire disc can be simply 
calculated by summing up the contributions of the different elements, i.e.

\begin {equation} 
\label{Tac}
\dot{M}(t) = \sum^{n}_{i=1}{\dot{M_i}(r,t)},
\end {equation}

\noindent where $n \equiv R_{\rm d}/\Delta r$ is the total number of surfaces, $R_{\rm d}$ is 
the adopted disc radius and $\Delta r$ is the width of the radial bins shown in Fig.~\ref{fig3}.

It is  important to emphasize that --inspired in the framework adopted by CEMs-- we are using 
cylindrical coordinates to check gas flows {\it through the disc surfaces}. 
A direct comparison with other gas accretion studies \citep[e.g.,][]{Peek2008,Joung2012,Nuza2014,Ceverino2015,Nelson2015} 
must be performed with care since gas infall is usually measured adopting spherical coordinates, 
i.e. no prefered direction is considered.

\subsection{Setting the cylinder's size}

To perform flux calculations, we adopt $R_{\rm d}=50\,$kpc and $z_{\rm d}=1\,$kpc for 
the radius and $z$-coordinate's range of the cylinder, respectively (see Fig.~\ref{fig3}).
The cylinder's radius is deliberately chosen larger than the optical radius of the galaxies
at $z=0$. We note that such a large $R_{\rm d}$ value is motivated by the fact
that gaseous discs are more extended than stellar ones; alowing us
to compute fluxes over their whole extension\footnote{The choice of $R_{\rm d}$ does not
  strongly affect our results for radially-integrated quantities
  because flux contributions at distances larger than optical radii of our simulated
  galaxies are significantly smaller than in central regions (see next section).}.
On the other hand, the adopted $z_{\rm d}$ gives a cylinder's height of $2\,$kpc
which is large enough to encompass the vertical distribution of simulated disc stars.
The influence of varying this parameter on the results is discussed in Appendix~\ref{zd}.

\begin{figure*}
\includegraphics[scale=0.35,angle=0]{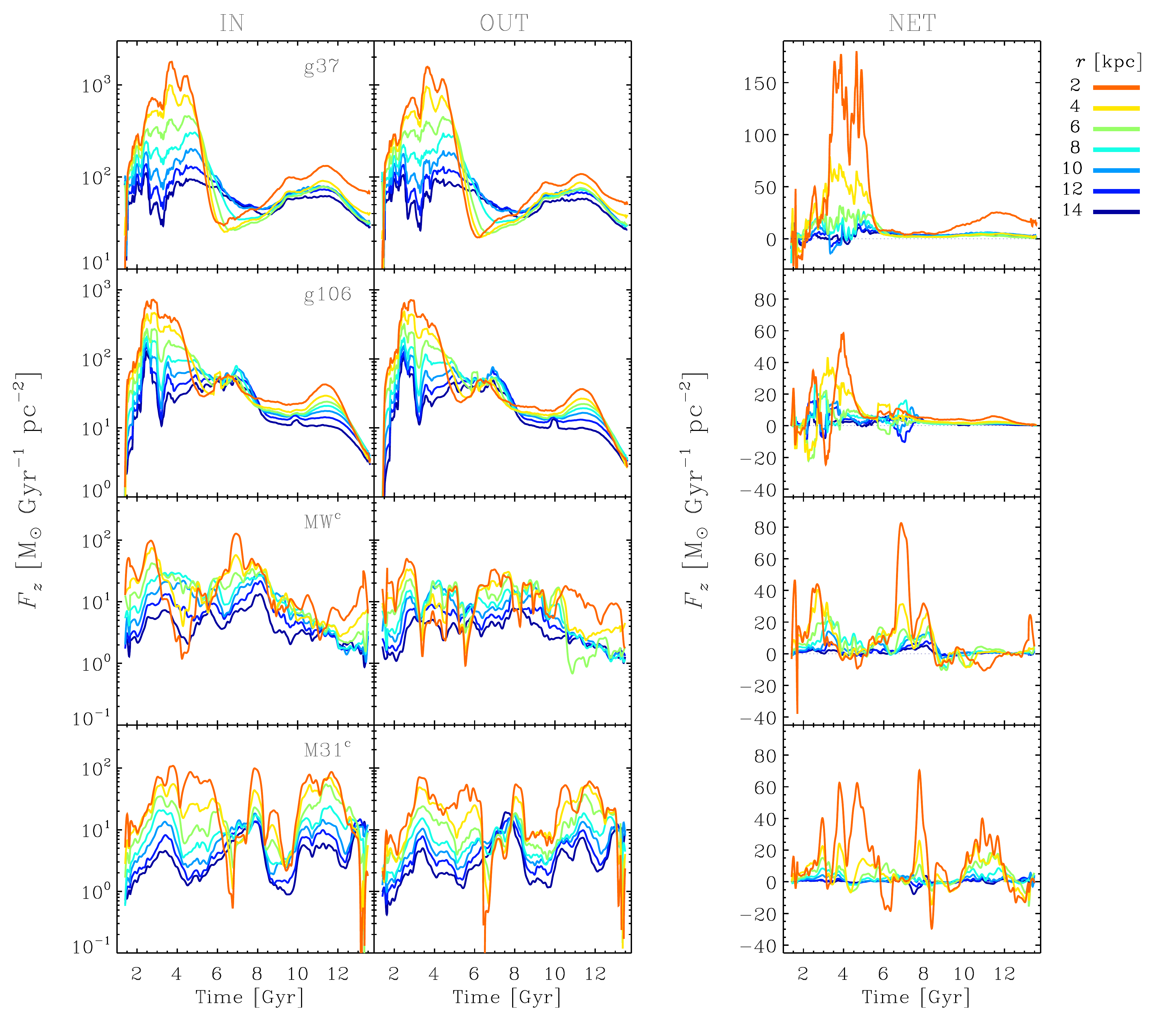}
\caption{Vertical infall and outflow fluxes (left-hand panels) and resulting net infall (right-hand panel) 
as a function of time at different radii for the simulated galaxies.}
\label{fig5}
\end{figure*}

\begin{figure*}
\includegraphics[scale=0.35,angle=0]{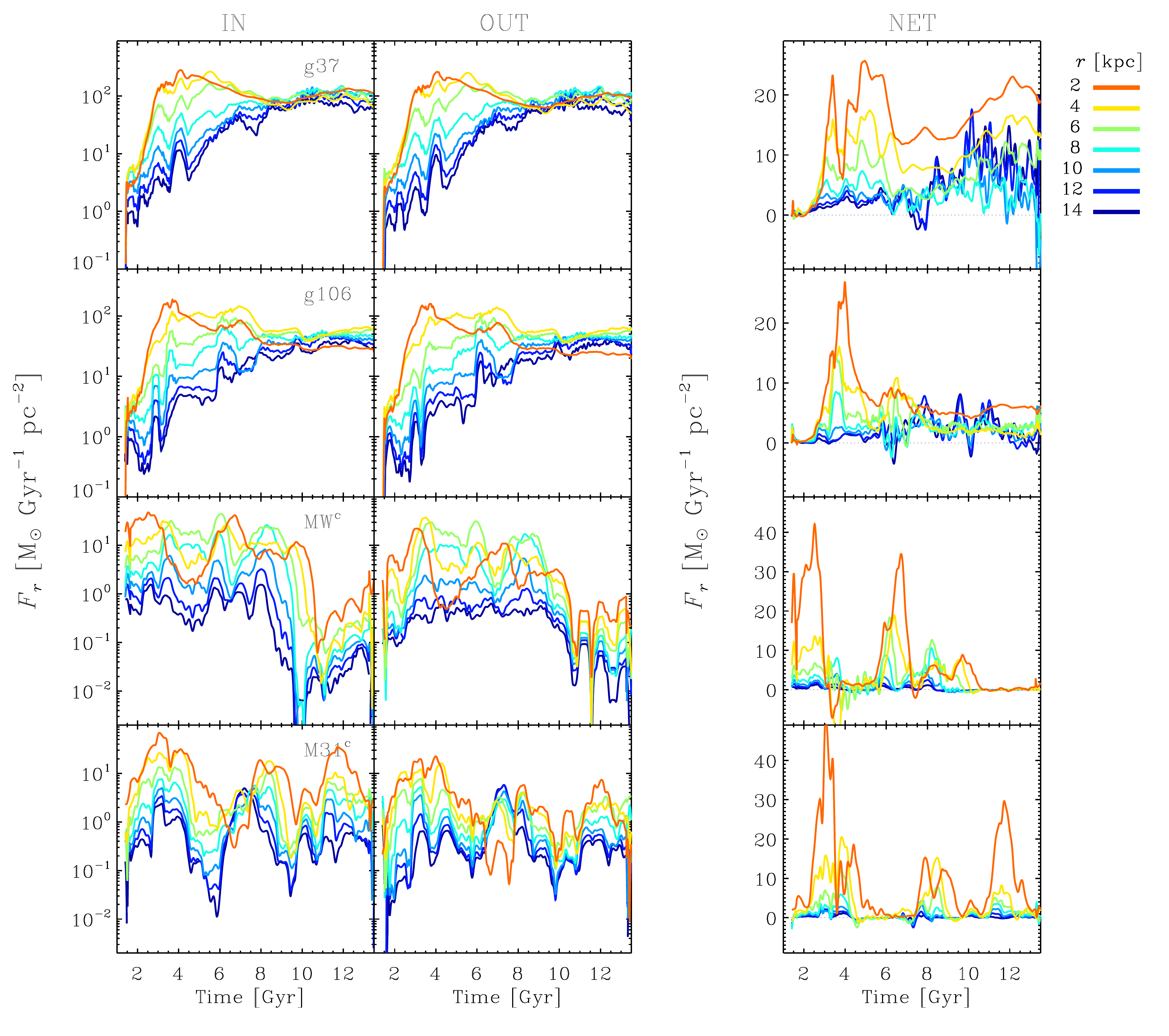}
\caption{Radial inward and outward fluxes (left-hand panels) and resulting net inward flux (right-hand panel) 
as a function of time at different radii for the simulated galaxies. 
}
\label{fig6}
\end{figure*}

\section{Results}
\label{res}

In this Section, we investigate the behaviour of the vertical and radial gas fluxes 
(inflowing and outflowing) as a function of time and radius for our four simulated
galaxies (Sections~\ref{VerticalRadial}~and~\ref{TimeInfall}) and gas accretion rates
focusing only on our MW candidates (Section~\ref{MWcandidates}). 
Note that, according to our definition, all fluxes are positive. In what follows, 
we define net flow rates as the difference between the infall and outflow fluxes, 
such that a positive value indicates a net gas infall.

\subsection{Vertical and radial gas fluxes vs. time and radius}
\label{VerticalRadial}

In order to have a first, qualitative view of the vertical and radial flow patterns 
of our simulated galaxies, we show, in Fig.~\ref{fig4}, the colour-coded logarithm of 
gas fluxes defined by Eqs.~\ref{flux_z} and \ref{flux_r}, 
where the horizontal and vertical axes indicate the cosmic time and disc radius, respectively. 
In this figure, colours represent flux values, normalized to the peak infall value of 
each galaxy, in order to identify the radii and times of the dominant gas flows. 
This procedure is performed separately for every flux component showing the vertical (radial) 
patterns in the left-hand (right-hand) panels.

The first column of Fig.~\ref{fig4} shows the vertical gas infall of our four simulated 
galaxies, indicating that gas inflows are higher both at early times 
--the first gigayears of evolution-- and at the innermost radii. 
At later times, the accretion is smoother, as well as extended in radius. 
This general behaviour is shared by all of our galaxies. However, it is clear that there are some 
qualitative differences between the two sets of simulations. In the CLUES galaxies, the accretion activity 
at late times is larger than for g37 and g106, as a result of the richer environment where they reside. 
Some of these accretion episodes are driven by `intermediate' 
(see e.g. the significant infall in MW$^{\rm c}$ at $t\sim 7\,$Gyr) and `major' mergers; in contrast 
to the `minor' ones taking place in g37 and g106 for most of their evolution (see Section~\ref{CLUES_gxs}). 
In particular, M31$^{\rm c}$ experiences several massive mergers after $z=1$, seen as infall 
peaks at $t\sim 8\,$Gyr and $t\gtrsim 10\,$Gyr, (see also Fig.~6 of \citealt{Scannapieco2015}) 
that disrupt the stellar disc and produce a bulge-dominated galaxy at the current epoch. 
Enhanced accretion in these simulations can also occur as a result of fountains 
triggered by feedback, as gas moves out from the disc owing to energy ejected during SN explosions 
to be reaccreted later.

\begin{figure*}
\includegraphics[scale=0.35,angle=0]{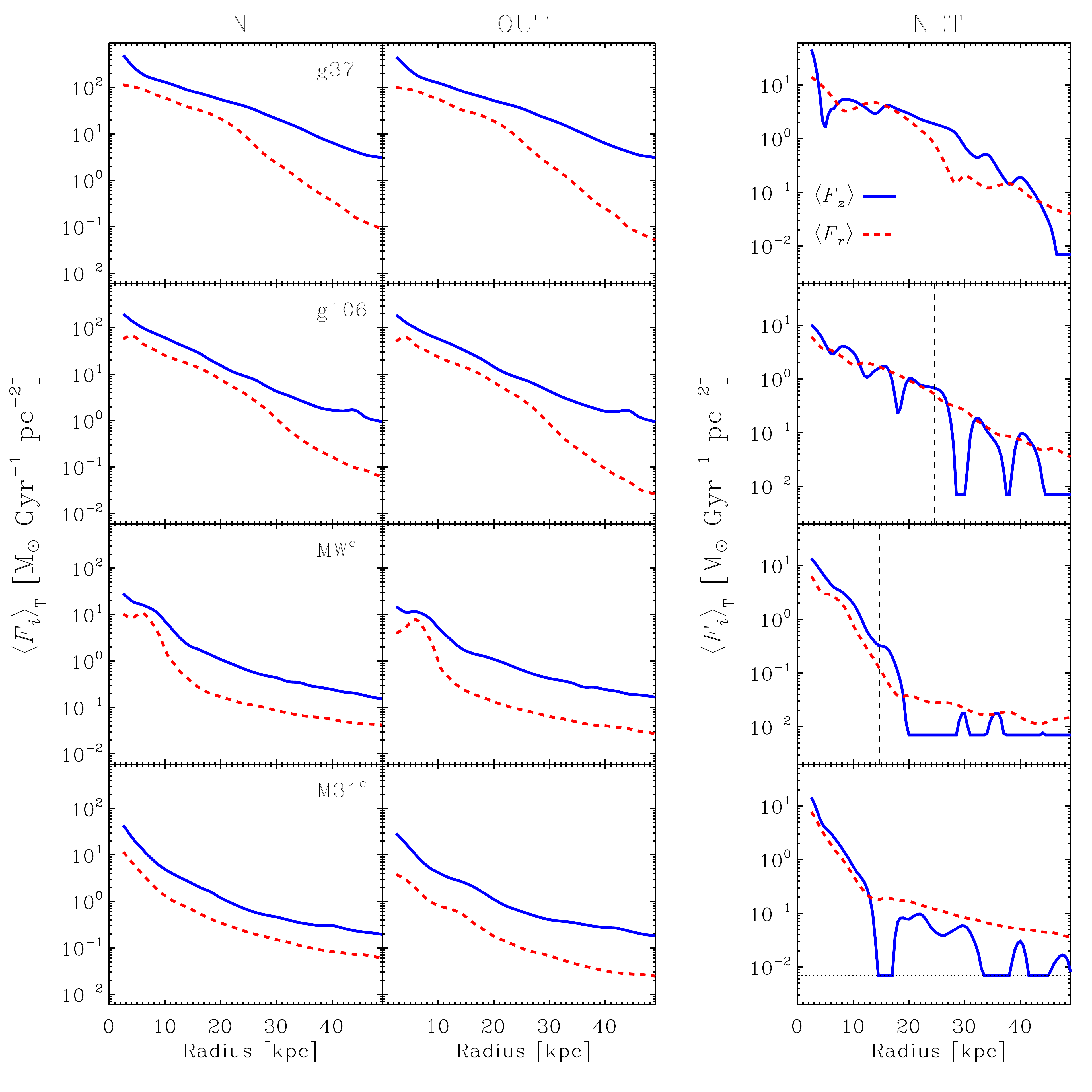}
\caption{Time-integrated infall and outflow fluxes (left-hand panels) and resulting 
net infall (right-hand panel) vs. radius for the simulated galaxies computed using 
Eq.~\ref{ir}. Both vertical and radial components are shown as solid 
and dashed lines respectively. Vertical dashed lines in the right-hand panels correspond
to galaxy optical radii.} 
\label{fig7}
\end{figure*}

\begin{figure*}
\includegraphics[scale=0.35,angle=0]{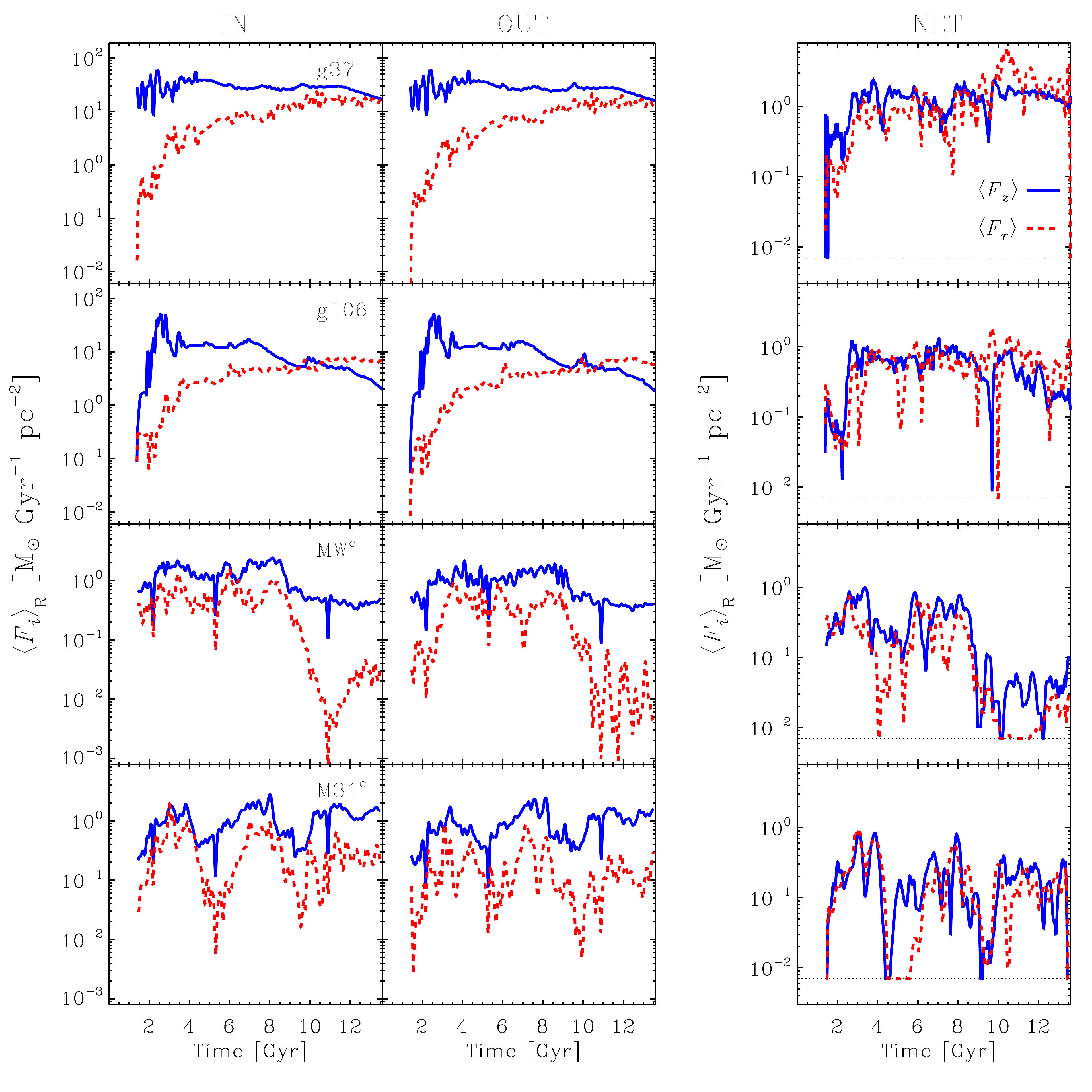}
\caption{Radially-averaged infall and outflow fluxes (left-hand panels) and resulting 
net infall (right-hand panel) vs. cosmic time for the simulated galaxies computed using 
Eq.~\ref{it}. Both vertical and radial components are shown as solid 
and dashed lines respectively.} 
\label{fig8}
\end{figure*}

The outflow of gas in the vertical direction is shown in the second column of Fig.~\ref{fig4}.  
Outflows can be generated both by SN explosions and/or by the passage of orbiting satellites. 
There is a significant correlation between inflow and outflow vertical patterns, possibly produced by 
the birth of stars in inflowing disc regions. This is more pronounced in the case of g37 and g106 
owing to the poor treatment of the hot gas phase in these galaxies that prevents 
the mixing of gas phases. Note that, in the simulations, the particularities of our two galaxy 
formation models can in fact lead to differences in the behaviour of the gas components. 
Ultimately, the rate of mass deposition in galaxy discs will depend on the interplay between 
gas inflows and outflows as it will be shown below.

The last two columns of Fig.~\ref{fig4} show the normalized inward and outward radial maps. 
Interestingly, radial and vertical flows are correlated. This is expected 
as gas falling on to the central galaxy regions will, in general, cross 
the boundaries of the cylinder in all directions. 
For the radial flows, we also detect higher gas fluxes at earlier times and small radii, although they are 
less concentrated in time compared to the vertical ones. Whereas in galaxies g37 and g106 the 
behaviour of the radial flows is smoother --slowly increasing in radius as time elapses-- for the 
CLUES galaxies the radial  accretion is more episodic; particularly at intermediate and late times. 
This is produced by the fact that, after the first infall event, accretion of material in g37 and g106 is 
less violent than in the CLUES simulation. As a result, the former are able to develope larger and 
denser gaseous discs at the present time.

In Figs.~\ref{fig5} and~\ref{fig6}, we quantitatively compare the gas flows for our simulated galaxies 
as a function of time for various radial bins. Besides the inflow and outflow patterns in each 
component (left-hand panels), we also show the net flux, 
i.e. $F_{i}(in)-F_{i}(out)$, such that a positive value indicates a pure gas inflow (right-hand panels). 
These figures indicate that inflow fluxes tend to be higher at early times and for the smallest radii.
This is most noticeable for g106 and MW$^{\rm c}$ galaxies, whose properties at $z=0$
better resemble those of the MW rather than g37 and M31$^{\rm c}$. A similar result
has been found by \cite{Courty2010} after measuring vertical inflows in a MW-size
simulated galaxy, although showing a somewhat stronger radial dependence at the present time. 
Quite generally, gas inflows are larger 
than outflows for most of the evolution. For all simulated galaxies, both the vertical and radial
net flows are dominated by the smallest radii, up to about $6\,$kpc from their centre. 
Additionally, vertical inflows at these radii are always dominant compared to radial flows; 
except for the very central regions at early times, where, although smaller, they are more similar. 
Radial flows remain small for most of the evolution but show a significant
net inward flux at early times in agreement with MCM14 (see their Fig.~12).
However, these are boosted during merger events at small and intermediate radial distances.
Note also that, in the case of g37, the radial inflow within $2\,$kpc is very pronounced 
and it is spread over a wide time interval, which is suggestive of the strong bar formed 
in this galaxy.

Galaxy g37 is the one with the highest vertical infall values at early times, followed by 
g106, MW$^{\rm c}$ and M31$^{\rm c}$, which might reflect the particular environment they inhabit 
and the halo mass variation in each galaxy formation model\footnote{Note that, in the case of CLUES 
galaxies, whereas MW$^{\rm c}$ is the least massive of the two at $z=0$, M31$^{\rm c}$ is the least 
massive at higher redshift (see \citealt{Scannapieco2015}).}. 
In the case of net radial flows, the situation is similar for all galaxies; 
although net flows at later times can be different, both in terms of their 
shape and magnitude. This is because, at late times, the formation history of a galaxy 
can significantly alter the accretion patterns, particularly in the presence of mergers, 
which are easily seen as flux enhancements. As already mentioned, this is fairly evident 
in CLUES but it can also be seen in g106 at $t\sim 7\,$Gyr, which corresponds to the only significant 
merger during its entire lifetime. Note that, even if satellite galaxies are not extremely 
gas-rich (as it is, in fact, the case of our CLUES simulations; see \citealt{Scannapieco2015}), 
and regardless of whether they fly-by or merge with the central galaxy, their interaction 
can drive gas inflows towards its center.

Our finding of higher accretion flux for earlier times and smaller radii is in agreement with 
the so-called `inside-out' disc formation scenario, in the sense that an early, strong infall leads to 
the formation of the inner region of the galaxy in a short timescale, followed by a subsequent, 
smoother growth of the gaseous disc, eventually allowing the formation of a stellar disc 
\citep[e.g.,][]{Larson1976,Brook2012}. We note, however, that the stellar component will not necessarily 
directly follow gas accretion history as, although star formation depends primarily on gas density, 
it is also influenced by feedback effects in a non-trivial manner. 
In this work we are mainly interested in studying gas inflow patterns on to the 
self-consistently evolved stellar disc that may eventually serve as fuel for star formation. 
Therefore, it is out of the scope of this paper to quantify how much of the accreted material 
is transformed into stars and ends up building the stellar disc.
\vfill

\begin{figure*}
\includegraphics[scale=0.37,angle=0]{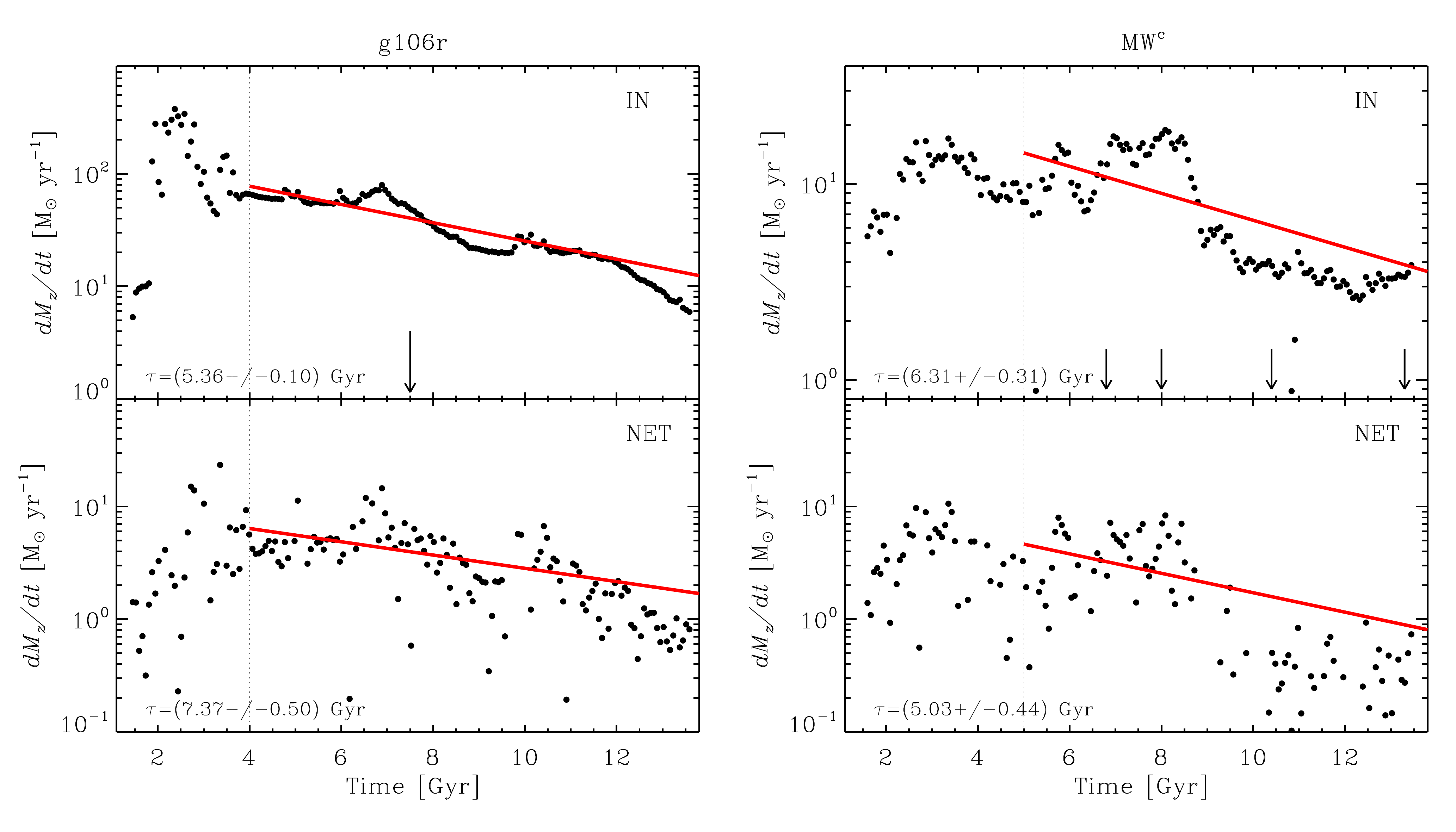}
\caption{Vertical infall (top panels) and net (bottom panels) gas accretion rate vs. cosmic time for 
our two MW galaxy candidates (see section~\ref{MWcandidates}). Solid lines are best fit exponentials: 
$\dot{M_i}(t)=A_i\exp^{-t/\tau_i}$, where $A_i$ is a normalization constant and $\tau_i$ is the 
decay timescale, for $t>t_{0}$. Best-fitting timescales and 
adopted $t_{0}$ (vertical dotted lines) values are shown. 
Arrows indicate the occurrence of significant merger events.} 
\label{fig9}
\end{figure*}

\subsection{Averaged fluxes}
\label{TimeInfall}

In this Section we focus on gas flows in the simulated galaxies studying separately their 
radial- and time-dependencies. As seen above, the picture may appear complex if we look at 
the flux as a function of time and radius, $F_i(r,t)$. In order to simplify this view, we integrate 
the infall (inward) and outflow (outward) fluxes in time and radius, producing two averaged 
flux functions that can be easily interpreted. After integrating in time and dividing by the 
total time spanned by the simulation, we obtain the following 
radial-dependent function

\begin{equation} 
  \label{ir}
  {\langle F_i \rangle}_{\rm T} \equiv \frac{\sum^{T}_{t=0} F_i(r,t) \Delta t}{T},
\end{equation}

\noindent where for each annulus of radius $r$ we sum the flux at each timestep 
$\Delta t$ over the entire time interval $T$. Thus, for a given radius, this function 
is proportional to the time-integrated mass per unit area in an arbitrary surface $S_i$. 
Similarly, for the radially-averaged temporal 
flux we can write

\begin {equation} 
\label{it}
{\langle F_i \rangle}_{\rm R} \equiv \frac{\sum^{n}_{i=1}{F_i(r,t)S_i}}{\sum^{n}_{i=1}{S_i}},
\end {equation}

\noindent which is simply the flux version of Eq.~\ref{Tac} divided by the 
corresponding total surface area. If cylinder's caps are considered the resulting 
function is proportional to the total mass accretion rate on to the disc plane at 
a given time. If, instead, radial surfaces are used, Eq.~\ref{it} provides a measure 
of the mass flow rate in the plane of the disc at a given cosmic time.

\begin{figure*}
\includegraphics[scale=0.4,angle=0]{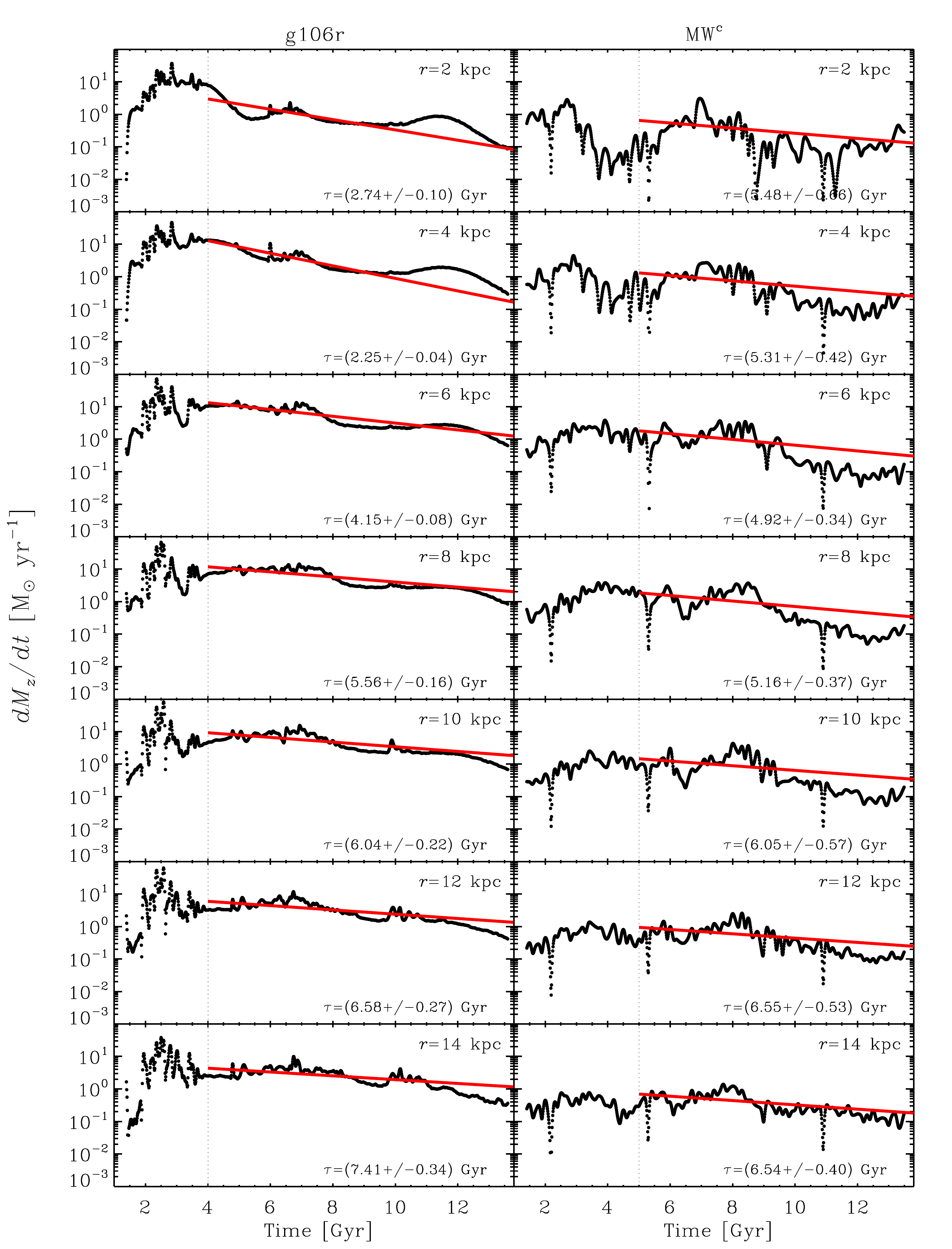}
\caption{Vertical infall rate vs. cosmic time split in radial bins for 
the g106r and MW$^{\rm c}$ simulated galaxies. Red solid lines show 
exponential fits obtained after $t_{0}=4,5\,$Gyr, respectively. Best-fitting 
timescales are also indicated.} 
\label{fig10}
\end{figure*}

\begin{figure*}
\includegraphics[scale=0.45,angle=0]{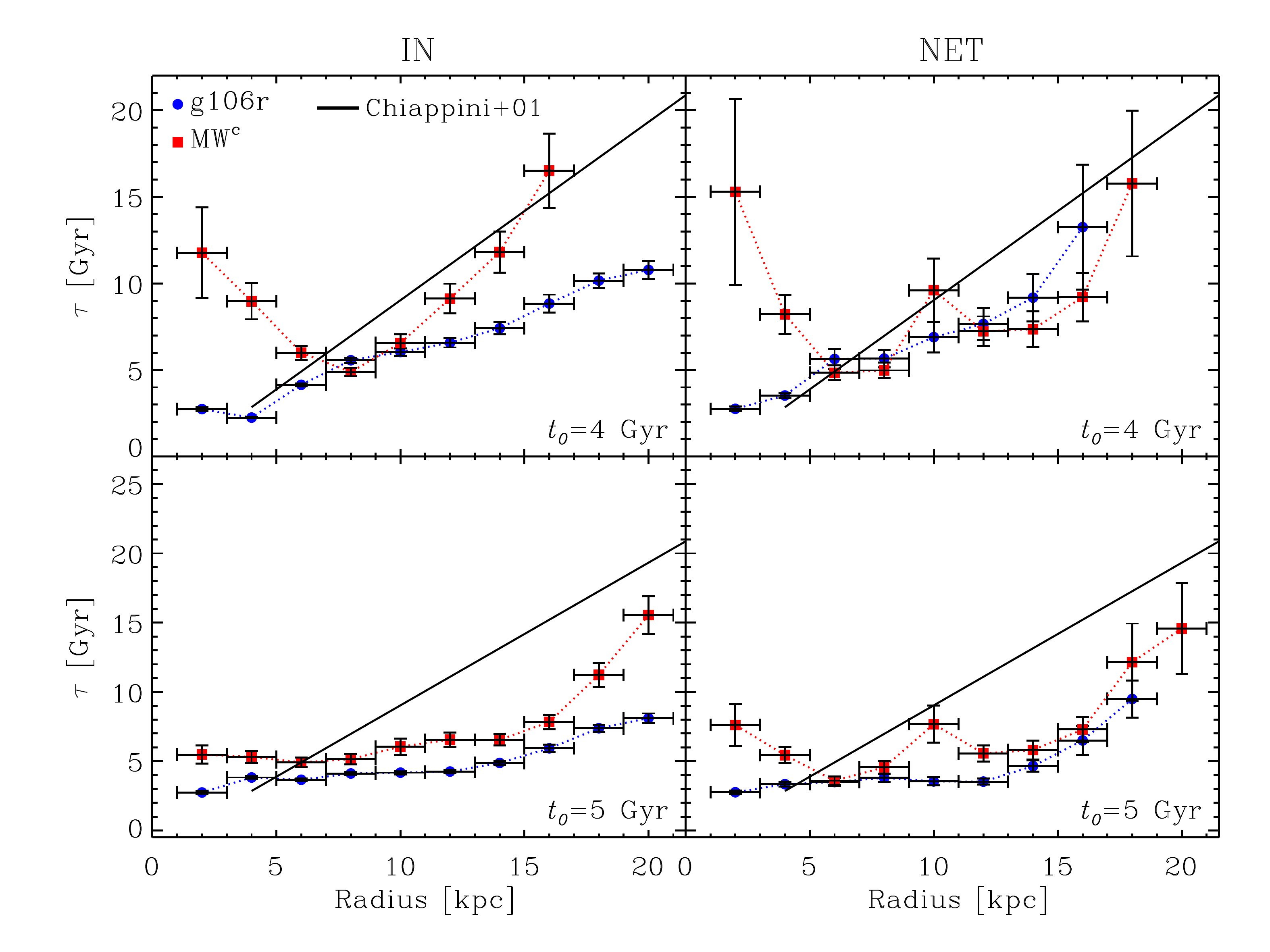}
\caption{Accretion decay timescale of gas vs. disc radius for our MW candidates
g106r and MW$^{\rm c}$ (solid symbols). Results obtained from the vertical infall 
(left-hand panels) and net accretion (right-hand panels) rates are shown.  
The minimum time, $t_{0}$, used to perform the fits is indicated in each panel. Data values can be found in Appendix~\ref{App_B}. For comparison, the linear approximation of Chiappini et al. (2001) for $r>4\,$kpc 
is also shown.} 
\label{fig11}
\end{figure*}

The left-hand panels of Fig.~\ref{fig7} show gas infall and outflow vs. radius 
as obtained from Eq.~\ref{ir} for the four simulated galaxies. As before, in the 
right-hand panel, we also plot the resulting net infall. Both vertical and radial 
mean flows are shown.
Decreasing fluxes with radius are consistent with the inside-out picture 
introduced in the previous section, showing that, during the whole evolution, 
more material is being accreted in the inner regions than in the outskirts. 
It is interesting to notice that, even beyond the optical radii
of the simulated galaxies ($r\gtrsim 15-35\,$kpc), gas is still being accreted 
in small amounts on to their gaseous discs, consistent to the findings of \cite{SLarsen2003}.
These plots demonstrate that gas infall accross the disc can be 
reasonably well described by an exponential law as also shown by \cite{Peek2008}. However, net vertical infall patterns complicate this picture, as they are typically noisier owing to the interplay between inflowing and outflowing 
material in certain disc regions. Interestingly, three out of four of our galaxies, show that net infall fluxes decrease significantly close to the edge of stellar discs (see vertical dashed lines in the right-hand panel of Fig.~\ref{fig7}), suggesting that star formation beyond these radii is less efficient. In this case, an exponential law up to the termination radius of the stellar discscan also be taken to describe the net infall of gas.

The existence of extended H\,{\sc i} discs in spirals is further supported by the
recent work of \cite{BlandHawthorn2017}. After computing the fraction of atomic
hydrogen in the presence of an external ionizing background,
these authors conclude that there is room for the detection of {\it proto-discs} of neutral
material at radii larger than previously thought. For instance, for column densities of $N_{\rm HI}=10^{18}\,$cm$^{-2}$,
the H\,{\sc i} discs might reach galactocentric distances of about $40-60\,$kpc, according to the different
scenarios considered. This is consistent with the flattened H\,{\sc i} structures
seen in our CLUES galaxies (see Fig.~10 of \citealt{Nuza2014}). From these simulations it is
possible to speculate on the origin of the accreting material at such large distances, which
is presumably associated to the circumgalactic halo.

The radial dependence of gas flows is similar in all galaxies although displaying different slopes. 
Radial inward and outward flux values are typically smaller than vertical ones, except for the 
very central regions, where they can be of the same order. In general, net radial and vertical 
fluxes show similar magnitudes. In terms of radial mass deposition within a given disc 
radius, $M_r(<r)$, this is mostly relevant for the central regions, i.e. at $r<4\,$kpc for our 
fiducial $z_{\rm d}$ value, being subdominant in comparison to net vertical mass inflows. 
In addition, there is a clear change of regime beyond the optical radius, with radial slopes 
getting shallower in three of our galaxies. This indicates that, beyond stellar discs, radial fluxes 
become roughly constant, eventually overtaking vertical ones at large radii.

The time dependence of the infall in galaxies has been shown to be much more complex. 
\cite{SLarsen2003} and \cite{Fraternali2012} found that the gas accretion history is in 
agreement with an exponential decay function. It is worth noting that these authors 
have reached the same conclusion after considering different accretion scenarios. For instance, 
\cite{SLarsen2003} assumed 
that the cooling of hot coronal gas is responsible for the gas accreted onto the galactic disc. 
However, this mechanism has been challenged by invoking thermal instability effects in the 
corona \citep{Binney2009,Nipoti2010}. On the other hand, \cite{Fraternali2012} obtained a smooth 
exponential accretion law by modelling the star formation rate (SFR) using 
the Kennicutt-Schmidt law, i.e. only considering gas actively involved 
in star formation, thus ignoring the bulk of the accretion. 
There is no unique explanation able to reproduce the accretion history of gas 
in an unbiased way. Moreover, several mechanisms are taking place 
simultaneously as shown by numerical simulations (e.g., \citealt{Nuza2014}),
where a halo mass transition of $\sim$$10^{12}\,$\Msun~separates the `cold' and `hot'
accretion regimes \citep{BirnboimDekel2003}\footnote{See, however, \cite{Nelson2015}
  for a recent discussion on the validity of a critical halo mass transition for the
  emergence of stable virial shocks.}.
We note that virial masses in
our simulated galaxies at $z=0$ lie just around this critical value and, therefore,
the two accretion modes could contribute. This is valid for our CLUES galaxies during most
of their evolution. In the case of g37 and g106, however, no hot halo gas is included
as a result of the adopted sticky-particle treatment of hydrodynamics (see Section~\ref{Martig_gxs}). 
In this work, we do not make any prior assumption about how gas is deposited on to the
stellar disc. We only use kinematic information provided by the hydrodynamical simulations to 
derive the infall flux, regardless of the mechanism responsible for the accretion. 
Therefore, within the limitations of the simulations, we believe that gas kinematics 
can be used to derive a coherent picture of the total accretion history of our MW-size 
objects in a reliable way. 

Fig.~\ref{fig8} shows the time dependence of gas fluxes computed using Eq.~\ref{it} for the 
four simulated galaxies. Both vertical and radial flux 
contributions are considered. The left-hand panel shows that radial fluxes are 
smaller than vertical ones during a significant fraction of the evolution 
(see also Figs.~\ref{fig5} and~\ref{fig6}). 
After first infall, net flows show the same tendency for all galaxies developing 
a significant stellar disc at the end of the simulation and with vertical-to-radial flow 
ratios of about 2. Interestingly, this tendency reverses for g106 at $t\gtrsim10\,$Gyr, 
i.e. well beyond the formation of the stellar disc in this simulation, when vertical 
accretion becomes negligible.  
The fact that radial fluxes show a net inward flow indicates that gas particles are 
able to migrate between different disc annuli. This is specially true at the smaller 
radii which are the ones dominating the signal. Nevertheless, as shown by MCM14 
for a rescaled version of g106 (see next section), most of the orbiting gas particles 
in the disc are non-migrators.

From these plots, it can also be seen that a simple 
exponential law can not explain the details of the whole gas accretion history in any of the 
galaxies under study. However, on average, an exponential law appears to be useful to describe 
fluxes over certain periods of time. This can be better applied after the first accretion episode, 
where average fluxes can be approximated using a linear function in the 
$\log F-t$ plane. In particular, the net infall evolution for galaxies g37, g106 and MW$^{\rm c}$ 
show a slight decline of just a few times for g37 to more than an order of 
magnitude for g106 and MW$^{\rm c}$. In the case of M31$^{\rm c}$, the very active 
merger history significantly flattens gas flows as a result of the continuous 
supply of material induced by mergers. It is worth 
recalling that M31$^{\rm c}$ does not have a dominant stellar disc at $z=0$, 
unlike actual M31, owing to the effects of late-time 
massive mergers \citep{Scannapieco2015}. 

\subsection{Gas accretion rate vs. cosmic time for MW candidates}
\label{MWcandidates}

In this section, we focus on the analysis of the MW galaxy candidates in 
our sample, i.e. g106 and MW$^{\rm c}$, to study in more detail several aspects 
of the gas infall process as a function of cosmic time.
Following MCM13, we downscaled the disc radius of g106 by a factor of $1.67$
and adjusted its rotational velocity at the solar radius to be $220\,$\kms.
This is done to obtain a model galaxy --dubbed g106r-- with dynamical and morphological
features closely matching the MW at the present time, which includes the central bar and
spiral arms. Other properties such as optical radius and gas-to-total-mass and B/T ratios
are also in good agreement with MW observations. We note that, by construction, g106 and its scaled
version share the same structural properties and, therefore, results obtained
with either version of the galaxy will be in qualitative agreement.
For MW$^{\rm c}$, only global MW disc properties are qualitatively reproduced, but in this case 
no fine-tuning is performed to the resulting galaxy output at the end of the simulation.

\subsubsection{Evidence for an exponential law}

One of the ingredients of CEMs for the MW is the parametrization of its gas accretion history. 
An early attempt to derive a cosmologically-motivated infall law for MW-like haloes
to test CEMs was performed by \cite{Colavitti2008}. These authors used dark matter-only
simulations to infer gas accretion from halo assembly history and concluded that the 
derived infall law was consistent to that of the `two-infall' model of \cite{Chiappini1997,Chiappini2001}.

In this work, we use our MW candidates as a proxy for gas evolution
in the Galaxy, thus considering both its cosmological and hydrodynamical
nature. We compute accretion rates in units of solar 
masses per year by multiplying gas fluxes by their corresponding surface areas.  
At late times, the vertical component accounts for most of 
the material coming from {\it outside} the disc, whereas radial fluxes mostly 
measures inward/outward flows {\it within} it. Therefore, the evolution of vertical 
accretion mainly quantifies the amount of material being incorporated into the system 
at any given time. 

Fig.~\ref{fig9} shows the vertical infall (top panels) and net (bottom panels) 
gas accretion rates vs. cosmic time for our two MW galaxy candidates. At early times, 
accretion rates start to increase reaching its maximum after a few gigayears and then decline towards 
net values within the range $0.6-1\,$\Msuny\;at the present day. We note that current rates 
measured at the virial radius of simulated galaxies with $M_{\rm vir}\sim 10^{12}\,$\Msun~(assuming spherical symmetry)
are of the order of $3-5$\Msuny\;\citep[e.g.,][]{Joung2012,Nuza2014}. If we assume that star-forming material 
accretes on to the disc at the typical rates found, the latter suggest that the 
efficiency of gas that is able to cool down and/or travel into the innermost regions of the 
Galaxy is $\sim$$10-30\%$. The top and bottom panels of Fig.~\ref{fig9} show very active accretion
at early times leading to a more gentle decline afterwards.
During this smoother accretion period stellar discs have the chance to steadily grow without suffering
significant perturbations, conversely to what happens in the early-infall episodes.
In the following, we define $t_{\rm 0}$ as the transition time between these two regimes; roughly corresponding
to the time where the crossing of dynamically-significant systems across the virial radius of the
galaxies have ceased (see e.g., Fig.~6 of \citealt{Scannapieco2015}).

As mentioned in the previous section, 
an exponential function alone is not able to describe the complexity of the {\it whole} gas accretion 
history of the galaxies. However, it can be safely taken as an average during certain 
periods of time (see solid lines).
This is specially evident after $t_0 \approx 4\,$Gyr for g106r.
For MW$^{\rm c}$, a similar behaviour is observed, although the more 
active merger history of this galaxy makes the evolution more complicated. An example of this is 
given by the two close `intermediate' mergers occurring at $t\sim6-8\,$Gyr, which have a 
sensible effect in the accretion curve (see blue arrows). 
In this galaxy, accretion rates start to systematically decline at somewhat longer times, 
i.e. after $t_0 \approx 5\,$Gyr.

The exponential fits shown in Fig.~\ref{fig9} indicate that 
the late-time accretion in our MW candidates takes places in timescales of 
$\tau\sim5-7\,$Gyr which, according to the fiducial disc's scale height adopted,
correspond to `thin-disc' structures in our simulations.
The latter can be considered `thin' in the sense that they are
formed by young, rotationally-supported stellar populations, in
contrast to simulated bulges \citep{Scannapieco2011,Martig2012}.
These timescales must be addressed in the context of CEMs
In these models, thin disc formation timescales are usually 
within a range of $5-7\,$Gyr, with associated accretion rate values of $0.4-1$\Msuny\;at the present 
day \citep{Chiappini2009,Kubryk2015}. These values are entirely consistent with our late-time
accretion results.

\subsubsection{Inside-out disc formation}
\label{inside_out}

As shown in previous sections, when integrating over the whole galaxy evolution, 
star-forming material is preferentially accreted in the central regions, eventually 
producing radially-growing discs. This `inside-out' formation of the Galaxy is modelled in some CEMs under 
simplifying assumptions; generally adopting a radially-increasing accretion timescale 
for the thin disc \citep[see e.g.][]{Romano2000,Chiappini1997,Chiappini2001,Chiappini2009}. 
In this way, it is possible to form stellar populations in the external regions of the 
disc later than in the inner parts. To study this further, we split the accretion rate evolution of 
our MW galaxy candidates in radial bins of width $2\,$kpc both for the vertical infall and net 
accretion components. 
Then, exponential fits are performed to compute the timescale as a function of radius. 
This procedure is illustrated in Fig.~\ref{fig10} for the vertical infall. 
All fits are done only for the late-time declining part of the accretion evolution 
as discussed in previous section. 
As expected, the resulting $\tau_D(R)$ timescales bracket the 
disc-averaged ones shown in the top panels of Fig.~\ref{fig9}.

In Fig.~\ref{fig11} we present one of the main results of our paper. This figure shows the decay 
timescale of the disc component of our MW candidates in the infall (left-hand panels) and net 
(right-hand panels) accretion components for different radial bins. Results using a minimum 
cosmic time of $t_{0}=4\,$Gyr (top panels) and $t_{0}=5\,$Gyr (bottom panels) 
are presented.
For completeness, we computed all decay timescales using either $t_{0}=4\,$Gyr (g106r) or
$5\,$Gyr (MW$^{\rm c}$), irrespectively of the galaxy being analysed (see Table~\ref{tab:app_b}). This can be taken to gauge the impact
of $t_{0}$ on the derived timescales. The linear approximation for the variation of $\tau_D(R)$ in the thin-disc 
postulated by \cite{Chiappini2001} in their CEM is also shown\footnote{This approximation is 
valid for galactocentric distances of $r>4\,$kpc. Note, however, that new constraints towards the inner 
Galaxy regions are now becoming available (\citealt{Chiappini2018}).}. Overall, we obtain similar trends between 
infall and net cases for the two galaxies under study. For $t_{0}=4\,$Gyr, g106r exhibits an 
increase with radius as expected for an inside-out formation scenario of the disc, although less pronounced 
than assumed in the CEM of \cite{Chiappini2001}. Interestingly, MW$^{\rm c}$ shows a downturn 
in the central regions, albeit inverting the trend at larger radii. This is somewhat 
expected since at $t\lesssim5\,$Gyr gas accretion in MW$^{\rm c}$ has not reached its maximum, 
thus flattening the evolution and increasing the corresponding timescales. 
If, instead, $t_{0}=5\,$Gyr is used, we get similar trends than obtained for g106r in the 
vertical infall and net cases.
Similarly, when $t_{0}=5\,$Gyr is adopted for g106r, original trends are also affected
as the resulting timescales are more sensitive to fluctuations in the evolution of gas accretion. 
When compared to the linear approximation of \cite{Chiappini2001}, results for
the decay timescales of MW$^{\rm c}$ towards larger galactocentric radii display less pronounced
slopes. Interestingly, some of these correlations show a curved shape that is reminiscent of the
one obtained in the analytical galaxy models of \cite{Molla2016} (see their Fig. 2).

From these plots, we can conclude that, during the more gentle, late-time decay phase of gas accretion, 
galaxy discs in our MW galaxy candidates are built in qualitative agreement with assumptions 
of CEMs of the Galaxy. However, this type of assembly strongly depends on the particular 
evolution history of the systems. A very active merger and/or gas accretion history can, in fact, 
significantly affect these trends. This is the case of M31$^{\rm c}$, where the growing stellar disc 
has been almost completely destroyed as a result of a late-time merger \citep{Scannapieco2015}. 
On the other hand, 
despite showing a quiet late-time evolution, the roughly constant accretion rate of g37 after 
$t=6\,$Gyr, produces a similar result; even in the presence of an already existing stellar 
disc. More details concerning these two galaxies are given in Appendix~\ref{App_B}.

\section{Summary and Discussion}
\label{concl}

We made an extensive analysis of gas fluxes on to the rotationally-supported stellar component 
of four different simulated galaxies focusing on two MW candidates, dubbed MW$^{\rm c}$ and g106r. 
The former is the MW galaxy candidate of \cite{Nuza2014}, whereas the latter is a rescaled 
version of the g106 galaxy of \cite{Martig2012} following MCM13 and MCM14. 
The simulations are consistent with a $\Lambda$CDM universe, 
use two different codes and setups and have been extensively tested to 
study the properties of galaxies similar to the MW. The different numerical techniques, 
SN feedback implementations and resolutions allow us to assess the robustness of the results.

The motivation for our work was to understand whether the assumptions 
of CEMs for gas infall patterns fueling SF in galaxy discs 
are consistent with the results of cosmological simulations that naturally 
include mass accretion, satellite interactions and mergers 
in an expanding universe. Nevertheless, these are not the only factors determining 
the SFR of a galaxy, as feedback effects are able to drive 
galactic winds and fountains affecting different regions of the disc 
in a unique way, both in terms of radius and height of the disc. 
Using the simulations, we obtained some general 
trends that are consistent with our current understanding of the MW 
evolution from CEMs, albeit not free from the complexities inherent 
to the physics of galaxy formation.
However, despite of the different setups and numerical techniques adopted,
similar conclusions are obtained from our set of simulated galaxies.

Our main results are discussed below:

\begin{itemize}
 \item All our galaxies show an early, high and rapidly decaying accretion phase --specially at 
  the smallest radii-- and a weaker, more gentle evolution at late times. We found that the latter 
  is specially important for the formation of a rotationally-supported galaxy stellar disc 
  as demonstrated by our simulations. 
  Our analysis shows that gas accretion is a complex process, greatly influenced 
  by details in the formation history of galaxies, which ultimately determine
  their fate and morphological type. Environmental effects could also play a role 
  (see e.g. \citealt{Creasey2015}); although a larger 
  sample of galaxies would be needed to properly quantify its importance 
  in MW-like systems.
  \\

  \item When time-integrating disc gas fluxes in the vertical and radial directions, 
  we found that infalling mass per unit area exponentially decays with radius as 
  already shown by \cite{Peek2008}. However, actual profiles often show 
  irregularities difficult to reproduce by a simple function, specially for net infall 
  patterns. In three of our four galaxies, net vertical infall profiles beyond 
  the optical radius decrease significantly suggesting that SF may be less efficient at those 
  radii. Conversely, radial profiles tend to flatten, indicating that most of the material being 
  accreted at large distances is mainly located within the plane of the rotating gaseous discs.
  \\
  
  \item The fact that a net inward flux is present in our simulations indicates that a fraction 
  of the available gas is able to migrate between different radial bins. According to MCM14 
  its fraction amounts to about $20-30\,\%$. This effect is more important 
  for the smallest disc radii, reaching its maximum at early times, and may have significant consequences 
  for CEMs neglecting radial flows at these times and locations. 
  At intermediate radial distances (including the solar neighborhood), we found small 
  --but nonzero-- net inward flows that might also have an impact 
  on galaxy metallicity gradients if not properly accounted for \citep[][]{Spitoni2011}.
  Moreover, during merger events, our simulations show that net radial flows are boosted by a
  significant amount, possibly affecting metallicity patterns at intermediate distances due to
  some degree of radial mixing. However, the analysis of chemical abundances, and their associated 
  temporal and spatial trends, is out of the scope of this paper and further work is needed 
  to conclude on these issues.  
  \\

 \item We found that a single exponential law cannot reproduce the complexities 
  of gas accretion history, although it can provide a reasonable approximation 
  to the mean during certain periods. This is specially evident at 
  late-times in our MW galaxy candidates, where infall patterns 
  decline with cosmic time. More importantly, postulating a dependence 
  on gas accretion with radius, such that the innermost galaxy regions present 
  higher accretion rates compared to the outskirts, appears to be a plausible assumption. 
  In particular, we found that accretion timescales in simulated disc-like 
  structures belonging to our MW candidates follow a similar trend than postulated in CEMs 
  (see e.g., \citealt{Romano2000,Chiappini2001,Chiappini2009,Molla2016}); although numerically 
  not entirely consistent.
  This behaviour goes in line with the so-called inside-out formation of 
  the thin-disc in the Galaxy. We stress, however, that this scenario may not 
  be valid for all galaxies as their specific merger and gas accretion histories 
  could have a strong impact on these trends, as it is shown in the case of 
  M31$^{\rm c}$ and g37 simulated galaxies (see Appendix~\ref{App_B}).

\end{itemize}

  On-going and future surveys will provide an enourmous amount of information 
  leading to a detailed description of properties in our Galaxy. This will allow 
  us to better trace its formation history, improve both chemical evolution and 
  galaxy formation models, and help us to better identify their joint strengths 
  and weaknesses.

\section*{Acknowledgments}

We thank the anonymous referee for valuable comments that helped to improve this paper.
SEN and CS are members of the Carrera del Investigador Cient\'{\i}fico 
of CONICET. SEN, CC and IM acknowledge support by the Deutsche Forschungsgemeinschaft 
under the grants NU 332/2-1, CH 1188/2-1 and MI 2009/1-1, respectively. 
CS acknowledges support from the Leibniz Gemeinschaft 
through grant SAW-2012-AIP-5 129. CC also acknowledges support from the 
ChETEC COST Action (CA16117), funded by COST (European Cooperation in 
Science and Technology). TCJ acknowledges support by Funda\c{c}\~ao CAPES 
and DAAD-CNPq-Brazil through a fellowship within the 
program ``Science without Borders''.



\bibliographystyle{mnras}
\bibliography{biblio} 

\begin{thebibliography}{}
\makeatletter
\relax
\def\mn@urlcharsother{\let\do\@makeother \do\$\do\&\do\#\do\^\do\_\do\%\do\~}
\def\mn@doi{\begingroup\mn@urlcharsother \@ifnextchar [ {\mn@doi@}
  {\mn@doi@[]}}
\def\mn@doi@[#1]#2{\def\@tempa{#1}\ifx\@tempa\@empty \href
  {http://dx.doi.org/#2} {doi:#2}\else \href {http://dx.doi.org/#2} {#1}\fi
  \endgroup}
\def\mn@eprint#1#2{\mn@eprint@#1:#2::\@nil}
\def\mn@eprint@arXiv#1{\href {http://arxiv.org/abs/#1} {{\tt arXiv:#1}}}
\def\mn@eprint@dblp#1{\href {http://dblp.uni-trier.de/rec/bibtex/#1.xml}
  {dblp:#1}}
\def\mn@eprint@#1:#2:#3:#4\@nil{\def\@tempa {#1}\def\@tempb {#2}\def\@tempc
  {#3}\ifx \@tempc \@empty \let \@tempc \@tempb \let \@tempb \@tempa \fi \ifx
  \@tempb \@empty \def\@tempb {arXiv}\fi \@ifundefined
  {mn@eprint@\@tempb}{\@tempb:\@tempc}{\expandafter \expandafter \csname
  mn@eprint@\@tempb\endcsname \expandafter{\@tempc}}}

\bibitem[\protect\citeauthoryear{{Anders} et~al.,}{{Anders}
  et~al.}{2014}]{Anders2014}
{Anders} F.,  et~al., 2014, \mn@doi [\aap] {10.1051/0004-6361/201323038}, \href
  {http://adsabs.harvard.edu/abs/2014A%26A...564A.115A} {564, A115}

\bibitem[\protect\citeauthoryear{{Anders} et~al.,}{{Anders}
  et~al.}{2017a}]{Anders2017a}
{Anders} F.,  et~al., 2017a, \mn@doi [\aap] {10.1051/0004-6361/201527204},
  \href {http://adsabs.harvard.edu/abs/2017A%26A...597A..30A} {597, A30}

\bibitem[\protect\citeauthoryear{{Anders} et~al.,}{{Anders}
  et~al.}{2017b}]{Anders2017b}
{Anders} F.,  et~al., 2017b, \mn@doi [\aap] {10.1051/0004-6361/201629363},
  \href {http://adsabs.harvard.edu/abs/2017A%26A...600A..70A} {600, A70}

\bibitem[\protect\citeauthoryear{{Aumer}, {White}  \& {Naab}}{{Aumer}
  et~al.}{2014}]{Aumer2014}
{Aumer} M.,  {White} S.~D.~M.,   {Naab} T.,  2014, \mn@doi [\mnras]
  {10.1093/mnras/stu818}, \href
  {http://adsabs.harvard.edu/abs/2014MNRAS.441.3679A} {441, 3679}

\bibitem[\protect\citeauthoryear{{Baglin}, {Michel}, {Auvergne}  \& {COROT
  Team}}{{Baglin} et~al.}{2006}]{Baglin2006}
{Baglin} A.,  {Michel} E.,  {Auvergne} M.,   {COROT Team} 2006, in Proceedings
  of SOHO 18/GONG 2006/HELAS I, Beyond the spherical Sun. p.~34

\bibitem[\protect\citeauthoryear{{Binney}, {Nipoti}  \& {Fraternali}}{{Binney}
  et~al.}{2009}]{Binney2009}
{Binney} J.,  {Nipoti} C.,   {Fraternali} F.,  2009, \mn@doi [\mnras]
  {10.1111/j.1365-2966.2009.15113.x}, \href
  {http://adsabs.harvard.edu/abs/2009MNRAS.397.1804B} {397, 1804}

\bibitem[\protect\citeauthoryear{{Birnboim} \& {Dekel}}{{Birnboim} \&
  {Dekel}}{2003}]{BirnboimDekel2003}
{Birnboim} Y.,  {Dekel} A.,  2003, \mn@doi [\mnras]
  {10.1046/j.1365-8711.2003.06955.x}, \href
  {http://adsabs.harvard.edu/abs/2003MNRAS.345..349B} {345, 349}

\bibitem[\protect\citeauthoryear{{Bland-Hawthorn}, {Maloney}, {Stephens},
  {Zovaro}  \& {Popping}}{{Bland-Hawthorn} et~al.}{2017}]{BlandHawthorn2017}
{Bland-Hawthorn} J.,  {Maloney} P.~R.,  {Stephens} A.,  {Zovaro} A.,
  {Popping} A.,  2017, \mn@doi [\apj] {10.3847/1538-4357/aa8f45}, \href
  {http://adsabs.harvard.edu/abs/2017ApJ...849...51B} {849, 51}

\bibitem[\protect\citeauthoryear{{Boeche} et~al.,}{{Boeche}
  et~al.}{2013}]{Boeche2013}
{Boeche} C.,  et~al., 2013, \mn@doi [\aap] {10.1051/0004-6361/201322085}, \href
  {http://adsabs.harvard.edu/abs/2013A%26A...559A..59B} {559, A59}

\bibitem[\protect\citeauthoryear{{Boeche} et~al.,}{{Boeche}
  et~al.}{2014}]{Boeche2014}
{Boeche} C.,  et~al., 2014, \mn@doi [\aap] {10.1051/0004-6361/201423974}, \href
  {http://adsabs.harvard.edu/abs/2014A%26A...568A..71B} {568, A71}

\bibitem[\protect\citeauthoryear{{Boissier} \& {Prantzos}}{{Boissier} \&
  {Prantzos}}{1999}]{Boissier1999}
{Boissier} S.,  {Prantzos} N.,  1999, \mn@doi [\mnras]
  {10.1046/j.1365-8711.1999.02699.x}, \href
  {http://adsabs.harvard.edu/abs/1999MNRAS.307..857B} {307, 857}

\bibitem[\protect\citeauthoryear{{Brook} et~al.,}{{Brook}
  et~al.}{2012}]{Brook2012}
{Brook} C.~B.,  et~al., 2012, \mn@doi [\mnras]
  {10.1111/j.1365-2966.2012.21738.x}, \href
  {http://adsabs.harvard.edu/abs/2012MNRAS.426..690B} {426, 690}

\bibitem[\protect\citeauthoryear{{Ceverino}, {Sanchez-Almeida},
  {Mu{\~n}oz-Tu{\~n}on}, {Dekel}, {Elmegreen}, {Elmegreen}  \&
  {Primack}}{{Ceverino} et~al.}{2015}]{Ceverino2015}
{Ceverino} D.,  {Sanchez-Almeida} J.,  {Mu{\~n}oz-Tu{\~n}on} C.,  {Dekel} A.,
  {Elmegreen} B.~G.,  {Elmegreen} D.~M.,   {Primack} J.,  2015, preprint, \href
  {http://adsabs.harvard.edu/abs/2015arXiv150902051C} {} (\mn@eprint {arXiv}
  {1509.02051})

\bibitem[\protect\citeauthoryear{{Chen}}{{Chen}}{2012}]{Chen2012}
{Chen} L.,  2012, in Nuclei in the Cosmos (NIC XII). p.~110

\bibitem[\protect\citeauthoryear{{Chiappini}}{{Chiappini}}{2009}]{Chiappini2009}
{Chiappini} C.,  2009, in {Andersen} J.,  {Nordstr{\"o}ara} {m} B.,
  {Bland-Hawthorn} J.,  eds,  IAU Symposium Vol. 254, IAU Symposium. pp
  191--196, \mn@doi{10.1017/S1743921308027580}

\bibitem[\protect\citeauthoryear{{Chiappini}, {Matteucci}  \&
  {Gratton}}{{Chiappini} et~al.}{1997}]{Chiappini1997}
{Chiappini} C.,  {Matteucci} F.,   {Gratton} R.,  1997, \apj, \href
  {http://adsabs.harvard.edu/abs/1997ApJ...477..765C} {477, 765}

\bibitem[\protect\citeauthoryear{{Chiappini}, {Matteucci}  \&
  {Romano}}{{Chiappini} et~al.}{2001}]{Chiappini2001}
{Chiappini} C.,  {Matteucci} F.,   {Romano} D.,  2001, \mn@doi [\apj]
  {10.1086/321427}, \href {http://adsabs.harvard.edu/abs/2001ApJ...554.1044C}
  {554, 1044}

\bibitem[\protect\citeauthoryear{{Chiappini}, {Minchev}, {Starkenburg}  \&
  {Valentini}}{{Chiappini} et~al.}{2018}]{Chiappini2018}
{Chiappini} C.,  {Minchev} I.,  {Starkenburg} E.,   {Valentini} M.,  eds, 2018,
  {Rediscovering our Galaxy}  IAU Symposium Vol. 334,
  \mn@doi{10.1017/S1743921318000789.
}

\bibitem[\protect\citeauthoryear{{Colavitti}, {Matteucci}  \&
  {Murante}}{{Colavitti} et~al.}{2008}]{Colavitti2008}
{Colavitti} E.,  {Matteucci} F.,   {Murante} G.,  2008, \mn@doi [\aap]
  {10.1051/0004-6361:200809413}, \href
  {http://adsabs.harvard.edu/abs/2008A%26A...483..401C} {483, 401}

\bibitem[\protect\citeauthoryear{{Courty}, {Gibson}  \& {Teyssier}}{{Courty}
  et~al.}{2010}]{Courty2010}
{Courty} S.,  {Gibson} B.~K.,   {Teyssier} R.,  2010, in {Debattista} V.~P.,
  {Popescu} C.~C.,  eds,  American Institute of Physics Conference Series Vol.
  1240, American Institute of Physics Conference Series. pp 131--134
  (\mn@eprint {arXiv} {1002.2383}), \mn@doi{10.1063/1.3458467}

\bibitem[\protect\citeauthoryear{{Creasey}, {Scannapieco}, {Nuza}, {Yepes},
  {Gottl{\"o}ber}  \& {Steinmetz}}{{Creasey} et~al.}{2015}]{Creasey2015}
{Creasey} P.,  {Scannapieco} C.,  {Nuza} S.~E.,  {Yepes} G.,  {Gottl{\"o}ber}
  S.,   {Steinmetz} M.,  2015, \mn@doi [\apjl] {10.1088/2041-8205/800/1/L4},
  \href {http://adsabs.harvard.edu/abs/2015ApJ...800L...4C} {800, L4}

\bibitem[\protect\citeauthoryear{{Dubois}, {Gavazzi}, {Peirani}  \&
  {Silk}}{{Dubois} et~al.}{2013}]{Dubois2013}
{Dubois} Y.,  {Gavazzi} R.,  {Peirani} S.,   {Silk} J.,  2013, \mn@doi [\mnras]
  {10.1093/mnras/stt997}, \href
  {http://adsabs.harvard.edu/abs/2013MNRAS.433.3297D} {433, 3297}

\bibitem[\protect\citeauthoryear{{Fenner} \& {Gibson}}{{Fenner} \&
  {Gibson}}{2003}]{Fenner2003}
{Fenner} Y.,  {Gibson} B.~K.,  2003, \mn@doi [\pasa] {10.1071/AS02047}, \href
  {http://adsabs.harvard.edu/abs/2003PASA...20..189F} {20, 189}

\bibitem[\protect\citeauthoryear{{Fraternali} \& {Tomassetti}}{{Fraternali} \&
  {Tomassetti}}{2012}]{Fraternali2012}
{Fraternali} F.,  {Tomassetti} M.,  2012, \mn@doi [\mnras]
  {10.1111/j.1365-2966.2012.21650.x}, \href
  {http://adsabs.harvard.edu/abs/2012MNRAS.426.2166F} {426, 2166}

\bibitem[\protect\citeauthoryear{{Genel} et~al.,}{{Genel}
  et~al.}{2014}]{Genel2014}
{Genel} S.,  et~al., 2014, \mn@doi [\mnras] {10.1093/mnras/stu1654}, \href
  {http://adsabs.harvard.edu/abs/2014MNRAS.445..175G} {445, 175}

\bibitem[\protect\citeauthoryear{{Gentile} et~al.,}{{Gentile}
  et~al.}{2013}]{Gentile2013}
{Gentile} G.,  et~al., 2013, \mn@doi [\aap] {10.1051/0004-6361/201321116},
  \href {http://adsabs.harvard.edu/abs/2013A%26A...554A.125G} {554, A125}

\bibitem[\protect\citeauthoryear{{Grand} et~al.,}{{Grand}
  et~al.}{2017}]{Grand2017}
{Grand} R.~J.~J.,  et~al., 2017, \mn@doi [\mnras] {10.1093/mnras/stx071}, \href
  {http://adsabs.harvard.edu/abs/2017MNRAS.467..179G} {467, 179}

\bibitem[\protect\citeauthoryear{{Guedes}, {Callegari}, {Madau}  \&
  {Mayer}}{{Guedes} et~al.}{2011}]{Guedes2011}
{Guedes} J.,  {Callegari} S.,  {Madau} P.,   {Mayer} L.,  2011, \mn@doi [\apj]
  {10.1088/0004-637X/742/2/76}, \href
  {http://adsabs.harvard.edu/abs/2011ApJ...742...76G} {742, 76}

\bibitem[\protect\citeauthoryear{{Hayden} et~al.,}{{Hayden}
  et~al.}{2014}]{Hayden2014}
{Hayden} M.~R.,  et~al., 2014, \mn@doi [\aj] {10.1088/0004-6256/147/5/116},
  \href {http://adsabs.harvard.edu/abs/2014AJ....147..116H} {147, 116}

\bibitem[\protect\citeauthoryear{{Hopkins}, {Kere{\v s}}, {O{\~n}orbe},
  {Faucher-Gigu{\`e}re}, {Quataert}, {Murray}  \& {Bullock}}{{Hopkins}
  et~al.}{2014}]{Hopkins2014}
{Hopkins} P.~F.,  {Kere{\v s}} D.,  {O{\~n}orbe} J.,  {Faucher-Gigu{\`e}re}
  C.-A.,  {Quataert} E.,  {Murray} N.,   {Bullock} J.~S.,  2014, \mn@doi
  [\mnras] {10.1093/mnras/stu1738}, \href
  {http://adsabs.harvard.edu/abs/2014MNRAS.445..581H} {445, 581}

\bibitem[\protect\citeauthoryear{{Hou} \& {Prantzos}}{{Hou} \&
  {Prantzos}}{2001}]{Hou2001}
{Hou} J.~L.,  {Prantzos} N.,  2001, \mn@doi [Nuclear Physics A]
  {10.1016/S0375-9474(01)00740-0}, \href
  {http://adsabs.harvard.edu/abs/2001NuPhA.688..411H} {688, 411}

\bibitem[\protect\citeauthoryear{{Hou}, {Chang}  \& {Chen}}{{Hou}
  et~al.}{2002}]{Hou2002}
{Hou} J.-L.,  {Chang} R.-X.,   {Chen} L.,  2002, \mn@doi [\cjaa]
  {10.1088/1009-9271/2/1/17}, \href
  {http://adsabs.harvard.edu/abs/2002ChJAA...2...17H} {2, 17}

\bibitem[\protect\citeauthoryear{{Joung}, {Putman}, {Bryan}, {Fern{\'a}ndez}
  \& {Peek}}{{Joung} et~al.}{2012}]{Joung2012}
{Joung} M.~R.,  {Putman} M.~E.,  {Bryan} G.~L.,  {Fern{\'a}ndez} X.,   {Peek}
  J.~E.~G.,  2012, \mn@doi [\apj] {10.1088/0004-637X/759/2/137}, \href
  {http://adsabs.harvard.edu/abs/2012ApJ...759..137J} {759, 137}

\bibitem[\protect\citeauthoryear{{Kubryk}, {Prantzos}  \&
  {Athanassoula}}{{Kubryk} et~al.}{2015}]{Kubryk2015}
{Kubryk} M.,  {Prantzos} N.,   {Athanassoula} E.,  2015, \mn@doi [\aap]
  {10.1051/0004-6361/201424171}, \href
  {http://adsabs.harvard.edu/abs/2015A%26A...580A.126K} {580, A126}

\bibitem[\protect\citeauthoryear{{Larson}}{{Larson}}{1976}]{Larson1976}
{Larson} R.~B.,  1976, \mnras, \href
  {http://adsabs.harvard.edu/abs/1976MNRAS.176...31L} {176, 31}

\bibitem[\protect\citeauthoryear{{Lehner} \& {Howk}}{{Lehner} \&
  {Howk}}{2011}]{Lehner2011}
{Lehner} N.,  {Howk} J.~C.,  2011, \mn@doi [Science] {10.1126/science.1209069},
  \href {http://adsabs.harvard.edu/abs/2011Sci...334..955L} {334, 955}

\bibitem[\protect\citeauthoryear{{Mackereth}, {Crain}, {Schiavon}, {Schaye},
  {Theuns}  \& {Schaller}}{{Mackereth} et~al.}{2018}]{Mackereth2018}
{Mackereth} J.~T.,  {Crain} R.~A.,  {Schiavon} R.~P.,  {Schaye} J.,  {Theuns}
  T.,   {Schaller} M.,  2018, preprint, \href
  {http://adsabs.harvard.edu/abs/2018arXiv180103593M} {} (\mn@eprint {arXiv}
  {1801.03593})

\bibitem[\protect\citeauthoryear{{Marinacci}, {Pakmor}  \&
  {Springel}}{{Marinacci} et~al.}{2014}]{Marinacci2014}
{Marinacci} F.,  {Pakmor} R.,   {Springel} V.,  2014, \mn@doi [\mnras]
  {10.1093/mnras/stt2003}, \href
  {http://adsabs.harvard.edu/abs/2014MNRAS.437.1750M} {437, 1750}

\bibitem[\protect\citeauthoryear{{Martig}, {Bournaud}, {Teyssier}  \&
  {Dekel}}{{Martig} et~al.}{2009}]{Martig2009}
{Martig} M.,  {Bournaud} F.,  {Teyssier} R.,   {Dekel} A.,  2009, \mn@doi
  [\apj] {10.1088/0004-637X/707/1/250}, \href
  {http://adsabs.harvard.edu/abs/2009ApJ...707..250M} {707, 250}

\bibitem[\protect\citeauthoryear{{Martig}, {Bournaud}, {Croton}, {Dekel}  \&
  {Teyssier}}{{Martig} et~al.}{2012}]{Martig2012}
{Martig} M.,  {Bournaud} F.,  {Croton} D.~J.,  {Dekel} A.,   {Teyssier} R.,
  2012, \mn@doi [\apj] {10.1088/0004-637X/756/1/26}, \href
  {http://adsabs.harvard.edu/abs/2012ApJ...756...26M} {756, 26}

\bibitem[\protect\citeauthoryear{{Martig}, {Minchev}  \& {Flynn}}{{Martig}
  et~al.}{2014}]{Martig2014}
{Martig} M.,  {Minchev} I.,   {Flynn} C.,  2014, \mn@doi [\mnras]
  {10.1093/mnras/stu1003}, \href
  {http://adsabs.harvard.edu/abs/2014MNRAS.442.2474M} {442, 2474}

\bibitem[\protect\citeauthoryear{{Matteucci}}{{Matteucci}}{2012}]{Matteucci2012}
{Matteucci} F.,  2012, {Chemical Evolution of Galaxies},
  \mn@doi{10.1007/978-3-642-22491-1.
}

\bibitem[\protect\citeauthoryear{{Matteucci} \& {Francois}}{{Matteucci} \&
  {Francois}}{1989}]{Matteucci1989}
{Matteucci} F.,  {Francois} P.,  1989, \mnras, \href
  {http://adsabs.harvard.edu/abs/1989MNRAS.239..885M} {239, 885}

\bibitem[\protect\citeauthoryear{{Miller} \& {Bregman}}{{Miller} \&
  {Bregman}}{2015}]{Miller2015}
{Miller} M.~J.,  {Bregman} J.~N.,  2015, \mn@doi [\apj]
  {10.1088/0004-637X/800/1/14}, \href
  {http://adsabs.harvard.edu/abs/2015ApJ...800...14M} {800, 14}

\bibitem[\protect\citeauthoryear{{Minchev}}{{Minchev}}{2015}]{Minchev2015}
{Minchev} I.,  2015, in Chemical and dynamical evolution of the Milky Way and
  Local Group. p.~14

\bibitem[\protect\citeauthoryear{{Minchev}, {Chiappini}  \& {Martig}}{{Minchev}
  et~al.}{2013}]{Minchev2013}
{Minchev} I.,  {Chiappini} C.,   {Martig} M.,  2013, \mn@doi [\aap]
  {10.1051/0004-6361/201220189}, \href
  {http://adsabs.harvard.edu/abs/2013A%26A...558A...9M} {558, A9}

\bibitem[\protect\citeauthoryear{{Minchev}, {Chiappini}  \& {Martig}}{{Minchev}
  et~al.}{2014}]{Minchev2014}
{Minchev} I.,  {Chiappini} C.,   {Martig} M.,  2014, \mn@doi [\aap]
  {10.1051/0004-6361/201423487}, \href
  {http://adsabs.harvard.edu/abs/2014A%26A...572A..92M} {572, A92}

\bibitem[\protect\citeauthoryear{{Minchev}, {Chiappini}  \& {Martig}}{{Minchev}
  et~al.}{2016}]{Minchev2016}
{Minchev} I.,  {Chiappini} C.,   {Martig} M.,  2016, \mn@doi [Astronomische
  Nachrichten] {10.1002/asna.201612404}, \href
  {http://adsabs.harvard.edu/abs/2016AN....337..944M} {337, 944}

\bibitem[\protect\citeauthoryear{{Moll{\'a}}, {D{\'{\i}}az}, {Gibson},
  {Cavichia}  \& {L{\'o}pez-S{\'a}nchez}}{{Moll{\'a}} et~al.}{2016}]{Molla2016}
{Moll{\'a}} M.,  {D{\'{\i}}az} {\'A}.~I.,  {Gibson} B.~K.,  {Cavichia} O.,
  {L{\'o}pez-S{\'a}nchez} {\'A}.-R.,  2016, \mn@doi [\mnras]
  {10.1093/mnras/stw1723}, \href
  {http://adsabs.harvard.edu/abs/2016MNRAS.462.1329M} {462, 1329}

\bibitem[\protect\citeauthoryear{{Moster}, {Somerville}, {Maulbetsch}, {van den
  Bosch}, {Macci{\`o}}, {Naab}  \& {Oser}}{{Moster} et~al.}{2010}]{Moster2010}
{Moster} B.~P.,  {Somerville} R.~S.,  {Maulbetsch} C.,  {van den Bosch} F.~C.,
  {Macci{\`o}} A.~V.,  {Naab} T.,   {Oser} L.,  2010, \mn@doi [\apj]
  {10.1088/0004-637X/710/2/903}, \href
  {http://adsabs.harvard.edu/abs/2010ApJ...710..903M} {710, 903}

\bibitem[\protect\citeauthoryear{{Naab} et~al.,}{{Naab}
  et~al.}{2014}]{Naab2014}
{Naab} T.,  et~al., 2014, \mn@doi [\mnras] {10.1093/mnras/stt1919}, \href
  {http://adsabs.harvard.edu/abs/2014MNRAS.444.3357N} {444, 3357}

\bibitem[\protect\citeauthoryear{{Nelson}, {Genel}, {Vogelsberger}, {Springel},
  {Sijacki}, {Torrey}  \& {Hernquist}}{{Nelson} et~al.}{2015}]{Nelson2015}
{Nelson} D.,  {Genel} S.,  {Vogelsberger} M.,  {Springel} V.,  {Sijacki} D.,
  {Torrey} P.,   {Hernquist} L.,  2015, \mn@doi [\mnras]
  {10.1093/mnras/stv017}, \href
  {http://adsabs.harvard.edu/abs/2015MNRAS.448...59N} {448, 59}

\bibitem[\protect\citeauthoryear{{Nelson}, {Genel}, {Pillepich},
  {Vogelsberger}, {Springel}  \& {Hernquist}}{{Nelson}
  et~al.}{2016}]{Nelson2016}
{Nelson} D.,  {Genel} S.,  {Pillepich} A.,  {Vogelsberger} M.,  {Springel} V.,
   {Hernquist} L.,  2016, \mn@doi [\mnras] {10.1093/mnras/stw1191}, \href
  {http://adsabs.harvard.edu/abs/2016MNRAS.460.2881N} {460, 2881}

\bibitem[\protect\citeauthoryear{{Nipoti}}{{Nipoti}}{2010}]{Nipoti2010}
{Nipoti} C.,  2010, \mn@doi [\mnras] {10.1111/j.1365-2966.2010.16705.x}, \href
  {http://adsabs.harvard.edu/abs/2010MNRAS.406..247N} {406, 247}

\bibitem[\protect\citeauthoryear{{Nuza}, {Parisi}, {Scannapieco}, {Richter},
  {Gottl{\"o}ber}  \& {Steinmetz}}{{Nuza} et~al.}{2014}]{Nuza2014}
{Nuza} S.~E.,  {Parisi} F.,  {Scannapieco} C.,  {Richter} P.,  {Gottl{\"o}ber}
  S.,   {Steinmetz} M.,  2014, \mn@doi [\mnras] {10.1093/mnras/stu643}, \href
  {http://adsabs.harvard.edu/abs/2014MNRAS.441.2593N} {441, 2593}

\bibitem[\protect\citeauthoryear{{Nuza}, {Chiappini}, {Scannapieco}, {Minchev},
  {Martig}  \& {Junqueira}}{{Nuza} et~al.}{2018}]{Nuza2018}
{Nuza} S.~E.,  {Chiappini} C.,  {Scannapieco} C.,  {Minchev} I.,  {Martig} M.,
   {Junqueira} T.~C.,  2018, in {Chiappini} C.,  {Minchev} I.,  {Starkenburg}
  E.,   {Valentini} M.,  eds,  IAU Symposium Vol. 334, Rediscovering Our
  Galaxy. pp 219--222 (\mn@eprint {arXiv} {1805.01588}),
  \mn@doi{10.1017/S1743921317008870}

\bibitem[\protect\citeauthoryear{{Pagel}}{{Pagel}}{2009}]{Pagel2009}
{Pagel} B.~E.~J.,  2009, {Nucleosynthesis and Chemical Evolution of Galaxies}

\bibitem[\protect\citeauthoryear{{Peek}, {Putman}  \& {Sommer-Larsen}}{{Peek}
  et~al.}{2008}]{Peek2008}
{Peek} J.~E.~G.,  {Putman} M.~E.,   {Sommer-Larsen} J.,  2008, \mn@doi [\apj]
  {10.1086/524374}, \href {http://adsabs.harvard.edu/abs/2008ApJ...674..227P}
  {674, 227}

\bibitem[\protect\citeauthoryear{{Portinari} \& {Chiosi}}{{Portinari} \&
  {Chiosi}}{1999}]{Portinari1999}
{Portinari} L.,  {Chiosi} C.,  1999, \aap, \href
  {http://adsabs.harvard.edu/abs/1999A%26A...350..827P} {350, 827}

\bibitem[\protect\citeauthoryear{{Prantzos} \& {Aubert}}{{Prantzos} \&
  {Aubert}}{1995}]{Prantzos1995}
{Prantzos} N.,  {Aubert} O.,  1995, \aap, \href
  {http://adsabs.harvard.edu/abs/1995A%26A...302...69P} {302, 69}

\bibitem[\protect\citeauthoryear{{Putman}, {Peek}  \& {Joung}}{{Putman}
  et~al.}{2012}]{Putman2012}
{Putman} M.~E.,  {Peek} J.~E.~G.,   {Joung} M.~R.,  2012, \mn@doi [\araa]
  {10.1146/annurev-astro-081811-125612}, \href
  {http://adsabs.harvard.edu/abs/2012ARA%26A..50..491P} {50, 491}

\bibitem[\protect\citeauthoryear{{Richter}}{{Richter}}{2012}]{Richter2012}
{Richter} P.,  2012, \mn@doi [\apj] {10.1088/0004-637X/750/2/165}, \href
  {http://adsabs.harvard.edu/abs/2012ApJ...750..165R} {750, 165}

\bibitem[\protect\citeauthoryear{{Richter} et~al.,}{{Richter}
  et~al.}{2017}]{Richter2017}
{Richter} P.,  et~al., 2017, \mn@doi [\aap] {10.1051/0004-6361/201630081},
  \href {http://adsabs.harvard.edu/abs/2017A%26A...607A..48R} {607, A48}

\bibitem[\protect\citeauthoryear{{Rocha-Pinto}, {Scalo}, {Maciel}  \&
  {Flynn}}{{Rocha-Pinto} et~al.}{2000}]{Rocha2000}
{Rocha-Pinto} H.~J.,  {Scalo} J.,  {Maciel} W.~J.,   {Flynn} C.,  2000, \aap,
  \href {http://adsabs.harvard.edu/abs/2000A%26A...358..869R} {358, 869}

\bibitem[\protect\citeauthoryear{{Romano}, {Matteucci}, {Salucci}  \&
  {Chiappini}}{{Romano} et~al.}{2000}]{Romano2000}
{Romano} D.,  {Matteucci} F.,  {Salucci} P.,   {Chiappini} C.,  2000, \mn@doi
  [\apj] {10.1086/309223}, \href
  {http://adsabs.harvard.edu/abs/2000ApJ...539..235R} {539, 235}

\bibitem[\protect\citeauthoryear{{Sancisi}, {Fraternali}, {Oosterloo}  \& {van
  der Hulst}}{{Sancisi} et~al.}{2008}]{Sancisi2008}
{Sancisi} R.,  {Fraternali} F.,  {Oosterloo} T.,   {van der Hulst} T.,  2008,
  \mn@doi [\aapr] {10.1007/s00159-008-0010-0}, \href
  {http://adsabs.harvard.edu/abs/2008A%26ARv..15..189S} {15, 189}

\bibitem[\protect\citeauthoryear{{Scannapieco}, {Tissera}, {White}  \&
  {Springel}}{{Scannapieco} et~al.}{2008}]{Scannapieco2008}
{Scannapieco} C.,  {Tissera} P.~B.,  {White} S.~D.~M.,   {Springel} V.,  2008,
  \mn@doi [\mnras] {10.1111/j.1365-2966.2008.13678.x}, \href
  {http://adsabs.harvard.edu/abs/2008MNRAS.389.1137S} {389, 1137}

\bibitem[\protect\citeauthoryear{{Scannapieco}, {White}, {Springel}  \&
  {Tissera}}{{Scannapieco} et~al.}{2011}]{Scannapieco2011}
{Scannapieco} C.,  {White} S.~D.~M.,  {Springel} V.,   {Tissera} P.~B.,  2011,
  \mn@doi [\mnras] {10.1111/j.1365-2966.2011.19027.x}, \href
  {http://adsabs.harvard.edu/abs/2011MNRAS.417..154S} {417, 154}

\bibitem[\protect\citeauthoryear{{Scannapieco} et~al.,}{{Scannapieco}
  et~al.}{2012}]{Scannapieco2012}
{Scannapieco} C.,  et~al., 2012, \mn@doi [\mnras]
  {10.1111/j.1365-2966.2012.20993.x}, \href
  {http://adsabs.harvard.edu/abs/2012MNRAS.423.1726S} {423, 1726}

\bibitem[\protect\citeauthoryear{{Scannapieco}, {Creasey}, {Nuza}, {Yepes},
  {Gottl{\"o}ber}  \& {Steinmetz}}{{Scannapieco}
  et~al.}{2015}]{Scannapieco2015}
{Scannapieco} C.,  {Creasey} P.,  {Nuza} S.~E.,  {Yepes} G.,  {Gottl{\"o}ber}
  S.,   {Steinmetz} M.,  2015, \mn@doi [\aap] {10.1051/0004-6361/201425494},
  \href {http://adsabs.harvard.edu/abs/2015A%26A...577A...3S} {577, A3}

\bibitem[\protect\citeauthoryear{{Singh}, {Majumdar}, {Nath}  \&
  {Silk}}{{Singh} et~al.}{2018}]{Singh2018}
{Singh} P.,  {Majumdar} S.,  {Nath} B.~B.,   {Silk} J.,  2018, \mn@doi [\mnras]
  {10.1093/mnras/sty1276}, \href
  {http://adsabs.harvard.edu/abs/2018MNRAS.478.2909S} {478, 2909}

\bibitem[\protect\citeauthoryear{{Sommer-Larsen}, {G{\"o}tz}  \&
  {Portinari}}{{Sommer-Larsen} et~al.}{2003}]{SLarsen2003}
{Sommer-Larsen} J.,  {G{\"o}tz} M.,   {Portinari} L.,  2003, \mn@doi [\apj]
  {10.1086/377685}, \href {http://adsabs.harvard.edu/abs/2003ApJ...596...47S}
  {596, 47}

\bibitem[\protect\citeauthoryear{{Spitoni} \& {Matteucci}}{{Spitoni} \&
  {Matteucci}}{2011}]{Spitoni2011}
{Spitoni} E.,  {Matteucci} F.,  2011, \mn@doi [\aap]
  {10.1051/0004-6361/201015749}, \href
  {http://adsabs.harvard.edu/abs/2011A%26A...531A..72S} {531, A72}

\bibitem[\protect\citeauthoryear{{Springel}}{{Springel}}{2005}]{Springel2005}
{Springel} V.,  2005, \mn@doi [\mnras] {10.1111/j.1365-2966.2005.09655.x},
  \href {http://adsabs.harvard.edu/abs/2005MNRAS.364.1105S} {364, 1105}

\bibitem[\protect\citeauthoryear{{Springel} et~al.,}{{Springel}
  et~al.}{2008}]{Springel2008}
{Springel} V.,  et~al., 2008, \mn@doi [\mnras]
  {10.1111/j.1365-2966.2008.14066.x}, \href
  {http://adsabs.harvard.edu/abs/2008MNRAS.391.1685S} {391, 1685}

\bibitem[\protect\citeauthoryear{{Stasi{\'n}ska} et~al.,}{{Stasi{\'n}ska}
  et~al.}{2012}]{Stasinska2012}
{Stasi{\'n}ska} G.,  et~al., eds, 2012, {Oxygen in the Universe}  EAS
  Publications Series Vol. 54

\bibitem[\protect\citeauthoryear{{Stinson}, {Brook}, {Macci{\`o}}, {Wadsley},
  {Quinn}  \& {Couchman}}{{Stinson} et~al.}{2013}]{Stinson2013}
{Stinson} G.~S.,  {Brook} C.,  {Macci{\`o}} A.~V.,  {Wadsley} J.,  {Quinn}
  T.~R.,   {Couchman} H.~M.~P.,  2013, \mn@doi [\mnras] {10.1093/mnras/sts028},
  \href {http://adsabs.harvard.edu/abs/2013MNRAS.428..129S} {428, 129}

\bibitem[\protect\citeauthoryear{{Teyssier}}{{Teyssier}}{2002}]{Teyssier2002}
{Teyssier} R.,  2002, \mn@doi [\aap] {10.1051/0004-6361:20011817}, \href
  {http://adsabs.harvard.edu/abs/2002A%26A...385..337T} {385, 337}

\bibitem[\protect\citeauthoryear{{Thilker}, {Braun}, {Walterbos}, {Corbelli},
  {Lockman}, {Murphy}  \& {Maddalena}}{{Thilker} et~al.}{2004}]{Thilker2004}
{Thilker} D.~A.,  {Braun} R.,  {Walterbos} R.~A.~M.,  {Corbelli} E.,  {Lockman}
  F.~J.,  {Murphy} E.,   {Maddalena} R.,  2004, \mn@doi [\apjl]
  {10.1086/381703}, \href {http://adsabs.harvard.edu/abs/2004ApJ...601L..39T}
  {601, L39}

\bibitem[\protect\citeauthoryear{{Tumlinson} et~al.,}{{Tumlinson}
  et~al.}{2013}]{Tumlinson2013}
{Tumlinson} J.,  et~al., 2013, \mn@doi [\apj] {10.1088/0004-637X/777/1/59},
  \href {http://adsabs.harvard.edu/abs/2013ApJ...777...59T} {777, 59}

\bibitem[\protect\citeauthoryear{{Walter}, {Brinks}, {de Blok}, {Bigiel},
  {Kennicutt}, {Thornley}  \& {Leroy}}{{Walter} et~al.}{2008}]{Walter2008}
{Walter} F.,  {Brinks} E.,  {de Blok} W.~J.~G.,  {Bigiel} F.,  {Kennicutt} Jr.
  R.~C.,  {Thornley} M.~D.,   {Leroy} A.,  2008, \mn@doi [\aj]
  {10.1088/0004-6256/136/6/2563}, \href
  {http://adsabs.harvard.edu/abs/2008AJ....136.2563W} {136, 2563}

\bibitem[\protect\citeauthoryear{{Wetzel}, {Hopkins}, {Kim},
  {Faucher-Gigu{\`e}re}, {Kere{\v s}}  \& {Quataert}}{{Wetzel}
  et~al.}{2016}]{Wetzel2016}
{Wetzel} A.~R.,  {Hopkins} P.~F.,  {Kim} J.-h.,  {Faucher-Gigu{\`e}re} C.-A.,
  {Kere{\v s}} D.,   {Quataert} E.,  2016, \mn@doi [\apjl]
  {10.3847/2041-8205/827/2/L23}, \href
  {http://adsabs.harvard.edu/abs/2016ApJ...827L..23W} {827, L23}

\bibitem[\protect\citeauthoryear{{Yepes}, {Gottl{\"o}ber}  \&
  {Hoffman}}{{Yepes} et~al.}{2014}]{Yepes2014}
{Yepes} G.,  {Gottl{\"o}ber} S.,   {Hoffman} Y.,  2014, \mn@doi [\nar]
  {10.1016/j.newar.2013.11.001}, \href
  {http://adsabs.harvard.edu/abs/2014NewAR..58....1Y} {58, 1}

\makeatother
\end{thebibliography}



\appendix

\section{Dependence of gas infall with height above the disc}
\label{zd}

To further investigate the dependence of infall rates with distance over 
the disc plane, we repeated our analysis for $z_{\rm d}=1,2,3\,$kpc, each 
value representing a total cylinder's height of $2,4,6\,$kpc, respectively.

\begin{figure*}
  \includegraphics[scale=0.38,angle=0]{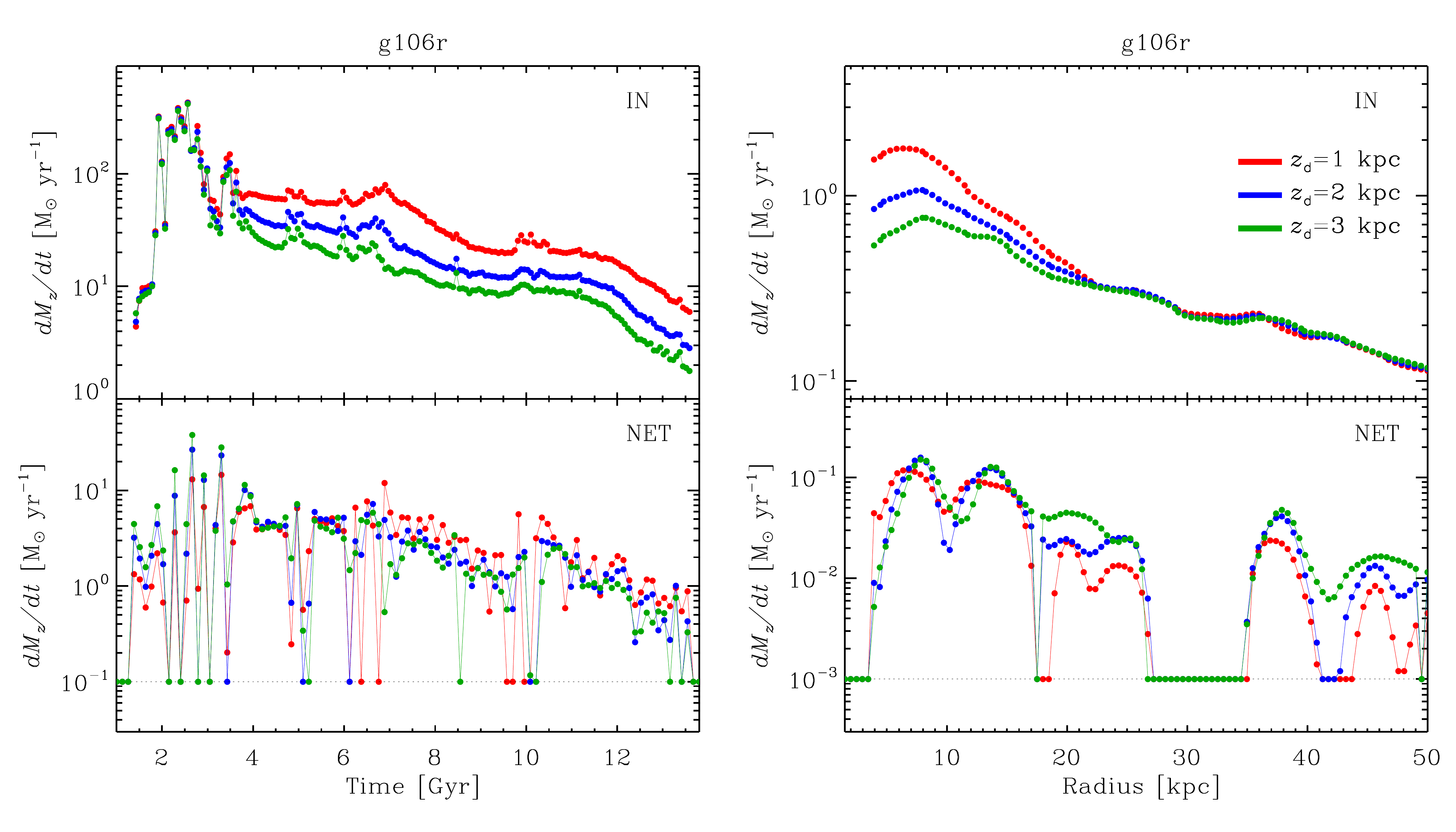}
  \includegraphics[scale=0.38,angle=0]{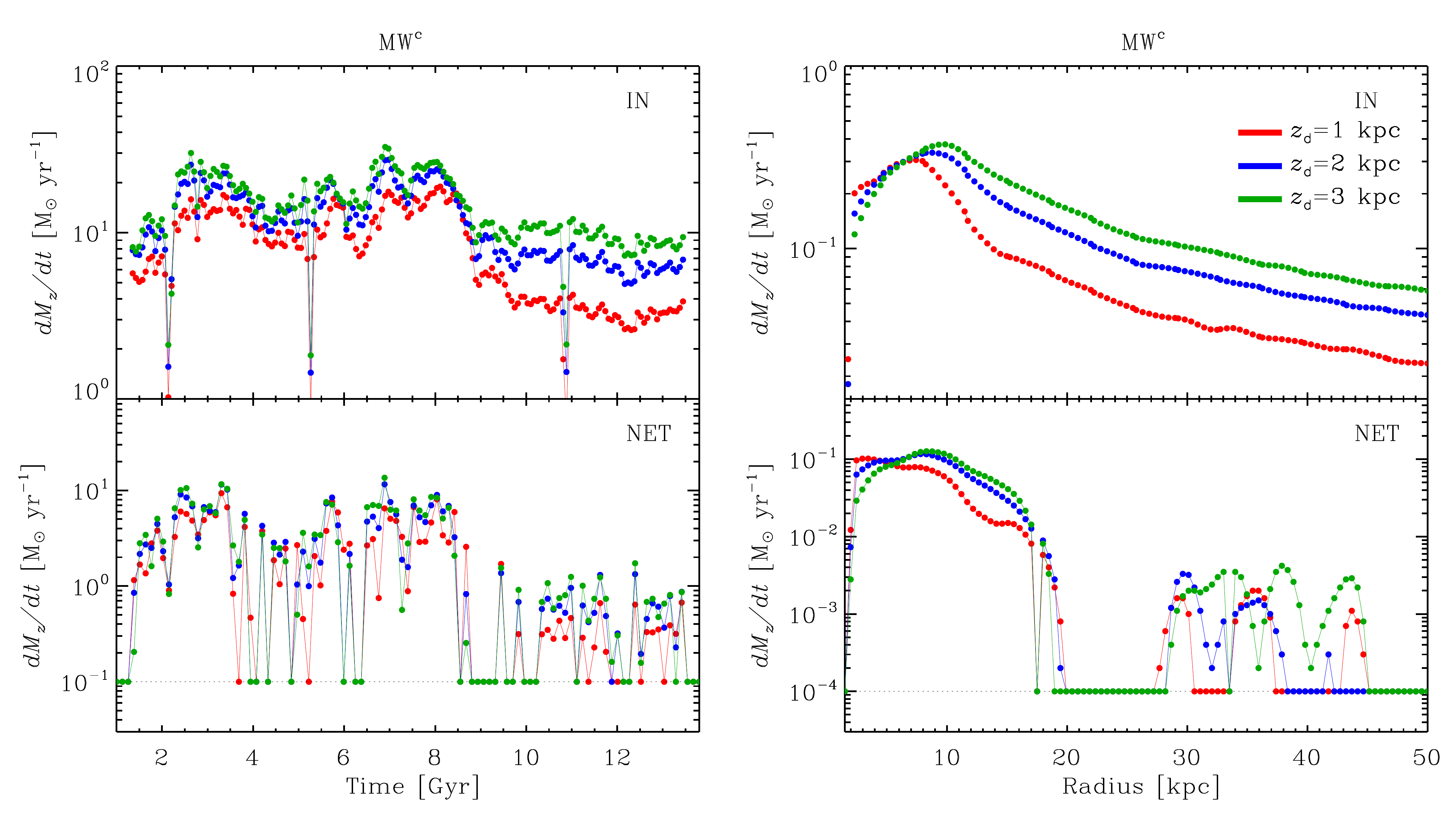}
    \caption{Vertical infall and net accretion rates vs. cosmic time (radially-averaged) and 
      disc radius (time-integrated) for the g106r (top panels) and MW$^{\rm c}$ (bottom panels) simulated galaxies
      computed at three different elevations above the disc plane.}
  \label{fig_appA}
\end{figure*}

Fig.~\ref{fig_appA} shows accretion rates as a function of time and radius
in the vertical infall and net cases for our two MW candidates. 
Whilst at early times less variation of the vertical infall for different heights
can be seen, it becomes more significant after a few gigayears, i.e. $t\gtrsim t_{0}$ (left-hand panels).
For g106r, the lowest accretion values correspond to the highest $z_{\rm d}$'s.
This occurs as a result of the poor treatment of the hot gas phase in this kind
of simulations, which prevents the formation of a hot gaseous halo around the galaxy
(see Section~\ref{Martig_gxs}). Therefore, most of the accreted material is
low entropy gas directly feeding the disc from filamentary-like structures. The lower the height
above the disc, the higher the fraction of colder (and clumpier) material. 
Conversely, this trend is inverted in the case of MW$^{\rm c}$. In this galaxy, the hot gaseous
halo is responsible for a significant fraction of the total accretion, typically increasing
with distance from the inner galaxy regions (see e.g., Fig. 14 of \citealt{Nuza2014}).

Right-hand panels show results obtained after integrating gas fluxes during the
entire galaxy lifetimes. For g106r, the vertical infall beyond the optical radius
does not depend on height as most of the material preferentially falls 
on to the central regions at higher rates. This is not the case of MW$^{\rm c}$,
which accretes material up to distances close to the virial radius owing to the presence
of the hot gaseous halo. Interestingly, for both galaxies, net vertical accretion splits into different curves
with the lowest accretion values corresponding to the lowest $z_{\rm d}$'s.
This simply reflects the fact that the intensity of galactic outflows always decreases as one moves
away from the disc plane, i.e. at higher elevations the net amount of accreted material is larger;
irrespectively of the nature of the dominant accretion modes in each galaxy.

In summary, although time and radial dependencies on accretion rates show 
some differences for different cylinder's heights, they can be easily interpreted 
and the resulting global trends are similar. We thus conclude that the particular choice of 
$z_{\rm d}=1\,$kpc adopted in this paper does not affect our main conclusions. 
However, caution must be taken when comparing with observations as absolute accretion 
values do vary with elevation above the disc midplane.

\section{Accretion timescales vs. disc radius in the general case}
\label{App_B}

\begin{figure*}
\includegraphics[scale=0.45,angle=0]{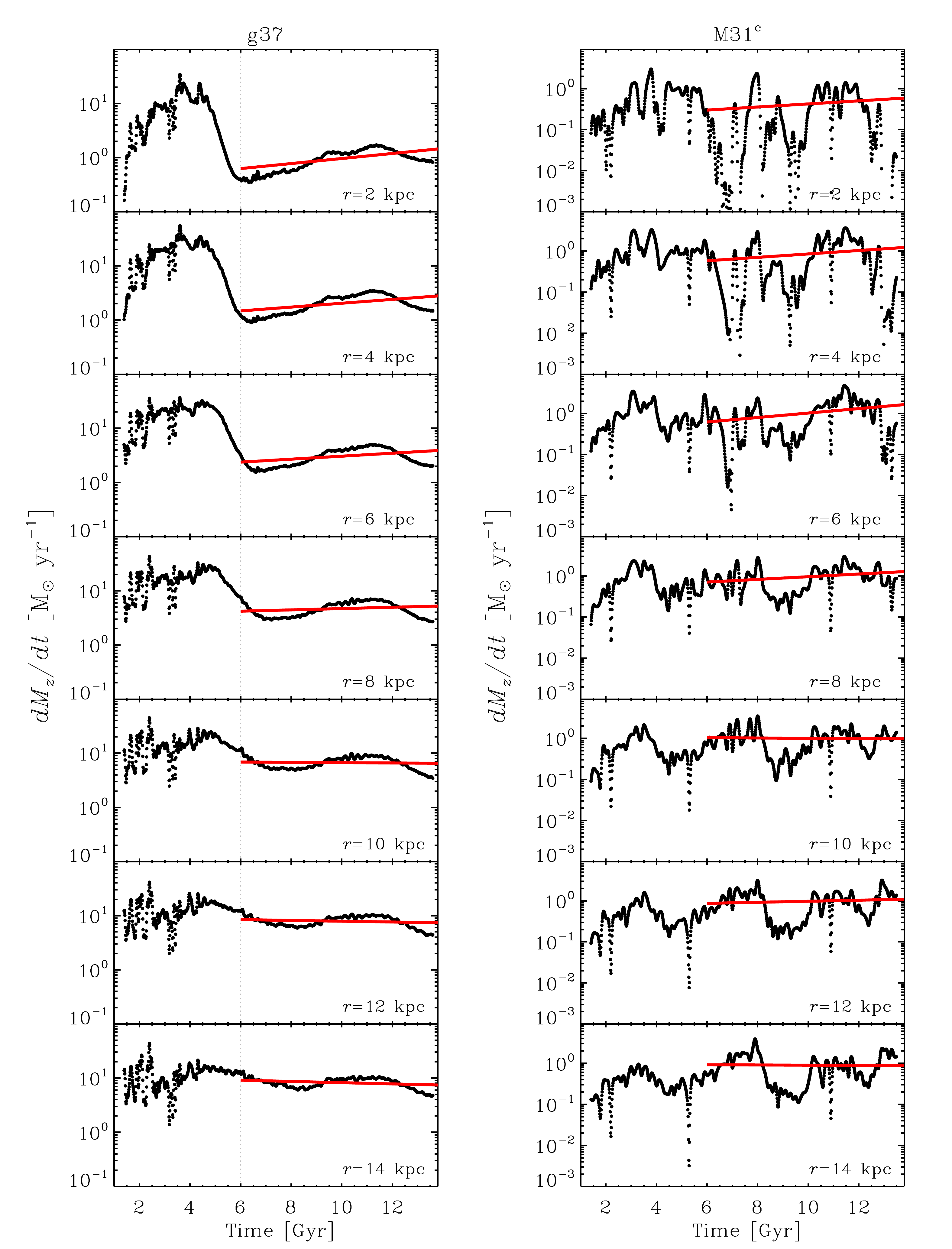}
\caption{Vertical infall rate vs. cosmic time split in radial bins for 
the g37 and M31$^{\rm c}$ simulated galaxies. Red solid lines show 
exponential fits obtained after $t_{0}=6\,$Gyr.}
\label{fig_appB}
\end{figure*}

As mentioned in Section~\ref{inside_out}, according to our simulations, 
the usual correlation 
between accretion timescale and radius invoked within the inside-out disc formation 
scenario is not straightforward. To illustrate this, 
in Fig.~\ref{fig_appB} we show the analogue of Fig.~\ref{fig10} for 
the case of the M31$^{\rm c}$ and g37 simulated galaxies. All fits are 
performed after the more rapid accretion period at early times, which roughly corresponds 
to $t>6\,$Gyr for these two galaxies. As seen in the plots, 
exponential fits in the $\log \dot{M}_z-t$ plane show a transition from 
positive slopes to flatter ones as a function of disc radius. 
This means that, although globally rates mildly decline 
or stay roughly constant with time, 
they actually {\it increase} exponentially during the late-time phase 
for the innermost radial bins. This is in sharp contrast to the behaviour 
observed in our MW candidates.
For a galaxy like M31$^{\rm c}$, which is undergoing an 
active merging phase that completely disrupts the existing 
stellar disc at late-times, this is expected. However, despite of having a very 
quiet merger history, g37 displays a similar behaviour; even in the presence 
of a stellar disc. 
This is mainly due to the formation of a strong bar at late-times 
as it can be seen, for instance, in the upper-right panel of Fig.~\ref{fig5} for $r\leq2\,$kpc. 
Moreover, at $t\gtrsim10\,$Gyr, material is also accreted at significant rates 
for the largest radii (see upper-left panel of Fig.~\ref{fig4}).  
As a result, these features significantly affect the way these specific discs 
form and evolve. 

\begin{table*}
\begin{center}
\caption{Accretion timescale, $\tau\pm\Delta\tau$, vs. galactocentric radius for the MW candidates shown in Fig.~\ref{fig11} (see Section~\ref{MWcandidates}). If no values are reported, no reliable fits were obtained at the corresponding radii.        
}
  \begin{tabular}{cc|cccc|cccc}
  \hline\hline
    & & \multicolumn{4}{c}{infall} & \multicolumn{4}{|c|}{net infall}\\
    \hline\hline
    & & \multicolumn{2}{|c|}{MW$^{\rm c}$} & \multicolumn{2}{|c|}{g106r} & \multicolumn{2}{|c|}{MW$^{\rm c}$} & \multicolumn{2}{|c|}{g106r} \\ 
    & $r$ & $\tau$ & $\Delta\tau$ & $\tau$ & $\Delta\tau$ & $\tau$ & $\Delta\tau$ & $\tau$ & $\Delta\tau$ \\
    & [kpc] & \multicolumn{2}{|c|}{[Gyr]}& \multicolumn{2}{|c|}{[Gyr]}& \multicolumn{2}{|c|}{[Gyr]}& \multicolumn{2}{|c|}{[Gyr]}\\
    \hline
    $t_0=4\,$Gyr & 2  & 11.77 & 2.62 & 2.74 & 0.10 & 15.29 & 5.37 & 2.76 & 0.13 \\
                 & 4  & 8.97  & 1.04 & 2.25 & 0.04 & 8.21  & 1.13  & 3.53 & 0.15 \\
                 & 6  & 6.00  & 0.39 & 4.15 & 0.08 & 4.85  & 0.42  & 6.64 & 0.58 \\
                 & 8  & 4.88  & 0.25 & 5.56 & 0.16 & 4.98  & 0.45  & 5.66 & 0.50 \\
                 & 10 & 6.56  & 0.51 & 6.04 & 0.22 & 9.60  & 1.83  & 6.90 & 0.89 \\
                 & 12 & 9.13  & 0.85 & 6.58 & 0.27 & 7.24  & 0.85  & 7.66 & 0.92 \\
                 & 14 & 11.81 & 1.20 & 7.41 & 0.34 & 7.35  & 1.03  & 9.18 & 1.38 \\
                 & 16 & 16.51 & 2.15 & 8.84 & 0.51 & 9.21  & 1.40  & 13.25 & 3.61 \\
                 & 18 &  --   &  -- & 10.16 & 0.42 & 15.77 & 4.21  & -- & -- \\
                 & 20 &  --   &  -- & 10.78 & 0.52 & --    & --    & -- & -- \\
        \hline
    $t_0=5\,$Gyr & 2  & 5.48  & 0.66 & 2.74 & 0.10 & 7.62  & 1.52 & 2.76 & 0.13 \\
                 & 4  & 5.31  & 0.42 & 3.82 & 0.10 & 5.44  & 0.57 & 3.34 & 0.18 \\
                 & 6  & 4.92  & 0.34 & 3.66 & 0.08 & 3.60  & 0.30 & 3.49 & 0.29 \\
                 & 8  & 5.16  & 0.37 & 4.11 & 0.10 & 4.56  & 0.47 & 3.80 & 0.30 \\
                 & 10 & 6.05  & 0.57 & 4.17 & 0.12 & 7.67  & 1.34 & 3.54 & 0.29 \\
                 & 12 & 6.55  & 0.53 & 4.25 & 0.12 & 5.56  & 0.59 & 3.53 & 0.22 \\
                 & 14 & 6.54  & 0.40 & 4.87 & 0.16 & 5.81  & 0.69 & 4.64 & 0.41 \\
                 & 16 & 7.83  & 0.52 & 5.93 & 0.26 & 7.29  & 0.91 & 6.50 & 1.04 \\
                 & 18 & 11.23 & 0.87 & 7.38 & 0.24 & 12.15 & 2.79 & 9.49 & 1.34 \\
                 & 20 & 15.54 & 1.35 & 8.10 & 0.34 & 14.57 & 3.29 & --   & -- \\
        \hline\hline
  
  \end{tabular}
  \label{tab:app_b}
\end{center}
\end{table*}


\end{document}